\documentclass[
nofootinbib,
reprint,
 amsmath,amssymb,
 nolongbibliography,
 aps,
 floatfix,
pra,
]{revtex4-2}

\usepackage{graphicx}
\usepackage{dcolumn}
\usepackage{bm}
\usepackage{cancel}
\usepackage{ulem}
\usepackage{amsmath}

\usepackage{xcolor}

\begin{document}

\preprint{APS/123-QED}

\title{Supermassive Binaries in Ultralight Dark Matter Solitons}

\author{Russell Boey${}^1$}
 \email{russell.boey@auckland.ac.nz}
 \author{Emily Kendall${}^1$}
 \email{emily-rose@carter-kendall.com}
\author{Yourong Wang${}^{1,2}$}
 \email{yourong.wang@uni-goettingen.de}
  \author{Richard Easther${}^{1}$}
 \email{r.easther@auckland.ac.nz}

\affiliation{${}^1$  Department of Physics,
University of Auckland,
Private Bag 92019,
Auckland, New Zealand}%
\affiliation{${}^2$  Institut f\"ur Astrophysik, Georg-August-Universit\"at G\"ottingen, D-37077 G\"ottingen, Germany}%

\date{\today}

\begin{abstract}
Ultralight (or fuzzy) dark matter (ULDM) is an alternative to cold dark matter. A key feature of ULDM is the presence of solitonic cores at the centers of collapsed halos. These would potentially increase the drag experienced by supermassive black hole (SMBH) binaries, changing their merger dynamics and the resulting  gravitational wave background.  We perform detailed simulations of high-mass SMBH binaries  in the soliton of a massive halo. We find more rapid decay than  previous simulations and semi-analytic approximations. We confirm expectations that the drag  depends strongly on the ULDM particle mass, finding masses greater than $10^{-21}$ eV could potentially alleviate the final parsec problem  and that ULDM may even suppress gravitational wave production at lower frequencies  in the pulsar timing band.

\end{abstract}

\maketitle


\section{\label{sec:intro}Introduction}

Ultralight dark matter (ULDM) is a dark matter candidate consisting of very light spin-0 particles \cite{Preskill:1982cy,Abbott:1982af,Hu:2000ke,Lesgourgues:2002hk,Suarez:2013iw,Graham:2015ouw,Marsh:2015xka,Ferreira:2020fam}, motivated by the low mass scalars which arise  in many string theoretic scenarios \cite{Svrcek:2006yi}. ULDM  masses between $10^{-23}$ and $10^{-20}$ eV may resolve some apparent discrepancies between the predictions of cold dark matter [CDM]  and observations on sub-galactic scales thanks to the wavelike nature of ULDM, which smooths small-scale structure relative to CDM. By contrast, on scales much larger than the de Broglie wavelength of the underlying particle ULDM behaves identically to CDM, making it an attractive dark matter candidate. 

In addition to their differing  substructures, ULDM halos have a central soliton, which may be particularly important to the dynamics of supermassive black hole [SMBH] binaries. Following a merger of their parent galaxies, dynamical friction causes the progenitor SMBH to migrate toward the center of the combined halo. Precise modeling shows that dynamical drag from gas \cite{Mayer:2007vk} can reduce the separation of the binary SMBH to a few parsecs in $\sim 10^8 \text{yr}$  \cite{Fang:2022cso}. This is  further reduced by the ejection of stars but this rapidly depletes the `loss cone', leading to the so-called `final parsec problem' where it apparently takes more than a Hubble time  to  reach separations from which gravitational radiation can  drive the actual  merger \cite{Milosavljevic:2002ht}.

In  simple models ULDM interacts purely gravitationally and can be treated non-relativistically. A Newtonian approach remains valid here because the dynamical friction experienced by SMBH is driven by the bulk motion of material rather than interactions near the event horizon and the ULDM is thus governed by the Schr\"{o}dinger-Poisson equations. It is well-established that progenitor ULDM halos merge to form a single halo comprising a central solitonic core enveloped by an outer Navarro-Frenk-White (NFW) profile \cite{Mocz:2017wlg, Kendall:2019fep,Zagorac:2022xic}. Following the  halo merger, the accompanying SMBH are expected to form a gravitationally bound binary within the resulting soliton. Because the formation timescale for the solitonic core is relatively short \cite{Paredes:2015wga} we consider a bound pair of SMBH moving inside a single soliton. While our focus here is on late-stage evolution of the SMBH within the soliton, previous work has indicated that stochastic density fluctuations in the outer NFW halo may disrupt the migration of the SMBH toward the central soliton \cite{Bar-Or:2018pxz, Dalal:2022rmp}. Our initial conditions therefore implicitly assume that this disruption is overcome.

Previous semi-analytic treatments extend Chandresekhar's  approach to  dynamical friction to ULDM \cite{Chandrasekhar:1943ys, Hui:2016ltb}. Comparisons of these semi-analytic predictions to numerical simulations have also been made in  specific regimes \cite{Lancaster:2019mde,Wang:2021udl}. However, it is clear that semi-analytic models cannot fully account for the range of dynamic effects exhibited by ULDM in numerical simulations. An excited soliton may transfer kinetic energy to perturbers within it \cite{Zagorac:2022xic, Boey:2024dks},  producing complex effects such as the `stone skipping' behavior of a single SMBH in a ULDM soliton \cite{Wang:2021udl}. Consequently, this paper focuses on highly-resolved numerical simulations that allow us to characterise both the drag and re-excitement experienced by SMBH binaries in ULDM solitons.

We use {\sc AxioNyx} \cite{Schwabe:2020eac},  an adaptive mesh refinement [AMR] enabled Schr\"{o}dinger-Poisson solver, which we extend to include black holes implemented as Plummer spheres. We initialise our simulations in an idealiced symmetric configuration, with two diametrically opposed SMBHs occupying circular orbits in an undisturbed soliton.  In contrast to the  stone-skipping  exhibited by a single black hole \cite{Wang:2021udl}, a symmetric binary undergoes an approximately power law decay in orbital radius, with a rapid, damped periodic modulation.

We perform simulations for a range of SMBH masses, and obtain numerical fits to the orbital decay curves. We compare these results to semi-analytic models \cite{Hui:2016ltb,Koo:2023gfm,Annulli:2020lyc} and previous simulations \cite{Koo:2023gfm} and find that the decay can be faster than previous analyses suggest. In particular, as the SMBH orbital radius decreases the soliton profile is `pinched' potentially increasing the central density by up to an order of magnitude and boosting the dynamical friction (see also Ref~\cite{Davies:2019wgi}). In addition, the decaying separation is overlaid by periodic modulation driven by `breathing' modes in the soliton excited by the black hole motion.    

For large SMBH moving in the soliton of a massive halo with  a ULDM particle mass at the higher end of its mass-range we see significant orbital decay even as the black hole separation falls below 1pc, suggesting that dynamical friction in a ULDM soliton could contribute to the solution of the `final parsec problem'. In addition it seems possible that ULDM could suppress gravitational wave emission at the low-frequency end of the pulsar timing band by imposing additional drag on binary SMBH. 
 
The structure of this paper is as follows: Section~\ref{sec:background} introduces the Schr\"{o}dinger-Poisson equations and describes their solution via {\sc AxioNyx} to model SMBH-ULDM interactions. Section~\ref{sec:Numerical} outlines the fiducial physical parameters for our simulations as well as choices of numerical parameters required for convergent results. Section \ref{sec:results} presents the results of our simulations. Both semi-analytic and purely empirical fits to the simulations are described in Section \ref{sec:semi-analytic}, and we discuss the implications of these results for the final parsec problem in Section~\ref{sec:parsec}. We conclude in Section \ref{sec:conclusion}.

\section{\label{sec:background}Background}

We work in a non-relativistic regime and consider the evolution of a pair of black holes in an initially circular orbit around the center of a soliton. ULDM is governed by the Schr\"{o}dinger-Poisson equations,
\begin{eqnarray}
    &i\hbar\dot{\psi} =-\frac{\hbar^2}{2m}\nabla^2\psi+m(\Phi_\mathrm{U}+\Phi_{\mathrm{BH,1}}+\Phi_{\mathrm{BH,2}})\psi&,
    \label{SP1} \\[10pt]
    &\nabla^2\Phi_\mathrm{U} =4\pi Gm|\psi|^2,&
    \label{SP2}
\end{eqnarray}
where $\psi$ is the ULDM wavefunction, $m$ is the ULDM particle mass, $\Phi_\mathrm{U}$ is the potential sourced by the ULDM and $\Phi_{\mathrm{BH,1}}$ and $\Phi_{\mathrm{BH,2}}$ are the contributions to the gravitational potential from the black holes. Each black hole is subject to a gravitational potential with contributions from the ULDM field and the other black hole, and evolves via
\begin{equation}
\ddot{\mathbf{x}}_{\mathrm{BH,{1/2}}}=-\nabla\Phi_\mathrm{U}(\mathbf{x}_{\mathrm{BH,{1/2}}})-\nabla\Phi_{\mathrm{BH,{2/1}}}(\mathbf{x}_{\mathrm{BH,{1/2}}}).
    \label{BH_evolution}
\end{equation}

In our simulations, the ULDM background is initialised as an isolated soliton. Simulations typically begin with the black holes begin inside the soliton core radius marking the soliton-NFW transition, around 3-4 core radii from the center \cite{Robles:2018fur,Kendall:2019fep}. The absence of the NFW halo in our simulations suppresses stochastic interactions between the soliton and the halo `granules' that can induce a random walk in the soliton position \cite{Schive:2019rrw, Chowdhury:2021zik}. Dynamical friction on binaries due to interactions with ULDM quasiparticles in the outer halo has previously been studied \cite{Bromley:2023yfi}, providing a complement to our simulations. 

The initial soliton profile is given by the spherically symmetric ground state solution to the governing Schr\"{o}dinger-Poisson equations in the absence of the black holes. This profile cannot be obtained analytically but is well-approximated by \cite{Schive:2014dra}
\begin{equation}   \rho(r)=1.9\frac{\left(\frac{m}{10^{-23}\text{eV}}\right)^{-2}\left(\frac{r_c}{\text{kpc}}\right)^{-4}}{\left[1+0.091\left(\frac{r}{r_c}\right)^2\right]^8}\textrm{M}_{\odot}\textrm{pc}^{-3},
\label{density_semi_analytic}
\end{equation}
where $m$ is the mass of the ULDM particle, $r$ is the radial distance from the center, and $r_c$ is the core radius, at which the density drops to half its peak value. We take this as the initial soliton profile in our simulations. This configuration is somewhat idealised, as the recent merger between the parent galaxies will give a soliton that has not  had time to relax to its ground state.

The dynamical friction experienced by the black hole due to ULDM was estimated in Ref.~\cite{Hui:2016ltb}
\begin{equation}
    F=\frac{4\pi G^2 M^2 \rho}{v^2} C,
    \label{eq:Hui_friction}
\end{equation}
where $M$ is the mass of the black hole and $v$ its velocity. $C$ is a coefficient of friction determined by the integral 
\begin{equation}
    C = \textrm{Cin}({2k\tilde{r}})+\frac{\sin(2k\tilde{r})}{2k\tilde{r}}-1 + \textrm{higher order terms},
    \label{Coefficient}
\end{equation}
where $k=m v/ \hbar$, and $\tilde{r}$ is a cutoff of integration that can be taken to represent the length of the wake caused by the black hole. This is nominally divergent unless $\tilde{r}$ is truncated at a finite distance, generally taken to be similar to be the orbital radius \cite{Wang:2021udl}. Following Koo {\it et al.}, we take this cutoff for a circular orbit of radius $r$ to be $\tilde{r} = \alpha r$, where $\alpha$ is an undetermined parameter somewhat less than unity \cite{Koo:2023gfm}. The resulting torque on a SMBH is 
\begin{equation}
\dot{L}=-Fr=-\frac{4\pi G^2 M^2 \rho}{v^2} Cr \, ,
\label{eq:torque}
\end{equation}
where $L$ is the angular momentum.   

Assuming an instantaneously circular orbit, for an equal mass binary
\begin{equation}
    v=\sqrt{\frac{GM_{enc}}{r}+\frac{GM}{4r}}
    \label{eq:circmotion}
\end{equation}
where $M_{enc}$ is the ULDM mass enclosed by the orbit. An expression for $\dot{r}$ in terms of $\dot{L}$ can be derived using $L=Mvr$,
\begin{align}
    \dot{L}&=\dot{r}\frac{dL}{dr}=\dot{r} \frac{M\sqrt{G}}{2} \times \nonumber\\&\left(\sqrt{\frac{M_{enc}+\frac{M}{4}}{r}}+\frac{dM_{enc}}{dr}\sqrt{\frac{r}{M_{enc}+\frac{M}{4}}}\right)\\&=\dot{r}\frac{M}{2}\left(v+\frac{4\pi r^2 \rho G}{v}\right),
    \label{eq:dLdtfull}
\end{align}
where we have assumed spherical symmetry to find ${dM_{enc}}/{dr}$. This can be combined with equation (\ref{eq:torque}) to give
\begin{equation}
    \dot{r}=-\frac{8\pi G^2 M \rho}{v^3+4\pi r^2 \rho Gv} Cr,
    \label{eq:rdotfull}
\end{equation}
a useful quantity for considering scaling relations.

\section{\label{sec:Numerical}Numerical Methods}

We use a modified version of {\sc AxioNyx} to simulate the black hole-soliton system. {\sc AxioNyx} is an AMR-enabled finite difference Schr\"{o}dinger-Poisson solver, providing high spatial and temporal resolution in the vicinity of the black holes, while dedicating less computational resource to regions which have a negligible effect on the evolution of the SMBH binary. We refine according to local ULDM density, and ensure that the black holes are inside the most refined region throughout the duration of the simulation.

The black holes are implemented as Plummer spheres, and their advancement proceeds according to Equation (\ref{BH_evolution}) via a fourth-order Runge-Kutta scheme (RK4). The value of $\Phi_\mathrm{U}$ is held fixed across the intermediate RK4 steps, and updated once at the end of each complete timestep. This quasi-static approximation of the ULDM potential is valid over the duration of each timestep since changes in the integrated ULDM potential occur on a longer timescale than those in the local density.

Interpolation over nearby grid points is used to find $\nabla\Phi_\mathrm{U}$ at the positions of the black holes. This interpolation is necessary since the black hole positions are continuous variables, while $\Phi_\mathrm{U}$ is only defined at the center of each grid cell. We determine the gradient of $\Phi_\mathrm{U}$ using the second-order central difference coefficients at the eight cells surrounding the black hole, and then perform a trilinear interpolation of these values to retrieve the gradient at the location of each black hole. The Plummer potential due to the black hole is
\begin{equation}
    \Phi_{BH}=-\frac{GM}{\sqrt{r^2+a^2}}\, ,
    \label{BH_potential}
\end{equation}
where $a$ is the Plummer radius, $\cal M$ is the mass of the black hole, and $r$ is the distance from the center of the black hole. The gradient of the Plummer potential is thus
\begin{equation}
    -\nabla\Phi_{BH}=-\frac{GM\textbf{r}}{(r^2+a^2)^{3/2}}.
\end{equation}

Equation (\ref{SP1}) is solved by separating into real and imaginary parts to retrieve
\begin{equation}
    \textrm{Re}(\dot{\psi})=-\frac{\hbar}{2m}\nabla^2\textrm{Im}(\psi)+\frac{m\Phi}{\hbar}\textrm{Im}(\psi)
\end{equation}
\begin{equation}
    \textrm{Im}(\dot{\psi})=\frac{\hbar}{2m}\nabla^2\textrm{Re}(\psi)-\frac{m\Phi}{\hbar}\textrm{Re}(\psi),
\end{equation}
where $\Phi$ is the sum of the ULDM and black hole potentials. The Laplacian is approximated using a standard 27-point stencil. The ULDM field is advanced using RK4 methods, again assuming a quasi-static approximation for the value of $\Phi_U$ across each timestep, but using the appropriate intermediate values of the black hole positions previously computed in their separate RK4 advancement. Equation (\ref{SP2}) is updated using a red-black Gauss-Seidel method, as described in Ref.~\cite{Almgren:2013sz}. 

\begin{figure}
    \centering
    \includegraphics[width=\linewidth]{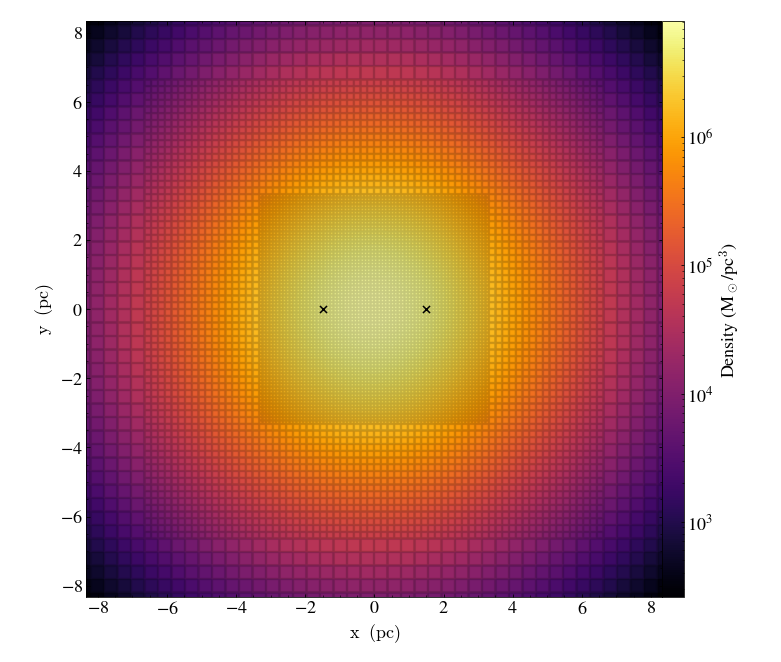}
    \caption{The initial density distribution and gridding in the central region of our simulation. The black holes are represented by crosses.}
    \label{fig:schematic}
\end{figure}

One subtlety is that when carrying out finite differencing operations for cells adjacent to a refinement boundary we need a suitable interpolation from coarse to fine grids. The {\sc AMReX} \cite{doi:10.1177/10943420211022811} framework underlying {\sc AxioNyx} offers a number of options and we use the conservative quartic interpolation on cell averaged data. This offered the best energy conservation and the closest match to a high resolution, unrefined soliton density profile in  testing. 

Our fiducial system consists of a soliton of mass $10^9 \textrm{M}_{\odot}$, with a ULDM particle mass of $10^{-21}$eV. The core radius is $\sim 2.2$pc and the central density is $\sim8\times 10^{6}\textrm{M}_{\odot}\textrm{pc}^{-3}$. The canonical core-halo relation \cite{Schive:2014dra,Zagorac:2022xic} implies an overall halo mass of  $3.8\times 10^{14} \textrm{M}_{\odot}$. 
Unless otherwise noted, each SMBH has a mass of $10^8 \textrm{M}_{\odot}$.

Our analysis thus focuses on representative examples rather than  an exhaustive scan of the relevant parameter space. However, this choice matches  Refs.~\cite{Broadhurst:2023tus,Koo:2023gfm} and corresponds to the estimated parameters of the binary system UGC4211 \cite{Koss:2023bvr}, facilitating comparisons with other work. The SMBH are  initially separated by $3$pc along the $x$ axis in a box of width 100~pc; {\sc AxioNyx} has periodic boundary conditions and a large box ensures that the soliton does not interact with its adjacent ``virtual'' counterparts. This initial separation is much smaller than that inferred for the astrophysical system UGC4211, reflecting our interest in the later stages of binary evolution. The black holes are taken to be in an initially circular orbit. Accounting for the contribution of the ULDM interior to the SMBH for our fiducial parameters the initial speeds are $\pm 584.14$kms$^{-1}$ in the $\pm\hat{y}$ directions. The default Plummer radius is $a=0.001$pc.

Typical runs use a default of 3 levels of refinement with a base grid of $128^3$ giving a resolution of $1024^3$ in the vicinity of the black holes. The initial refinement boundaries are at 12.5pc, 6.25pc, and 3.125pc from the center of the grid in each dimension but  the boundaries change as the density of the soliton evolves - a requirement for numerical stability. The default timestep in {\sc AxioNyx} is determined by the Courant-Friedrichs-Lewy (CFL) condition, as described in \cite{Almgren:2013sz}. The value of $\sigma^{\textrm{CFL,hyp}}$ is 0.8 in our simulations. The initial configuration is illustrated in Figure~\ref{fig:schematic}.

\begin{figure}[tb]
    \centering
    \includegraphics[width=\linewidth]{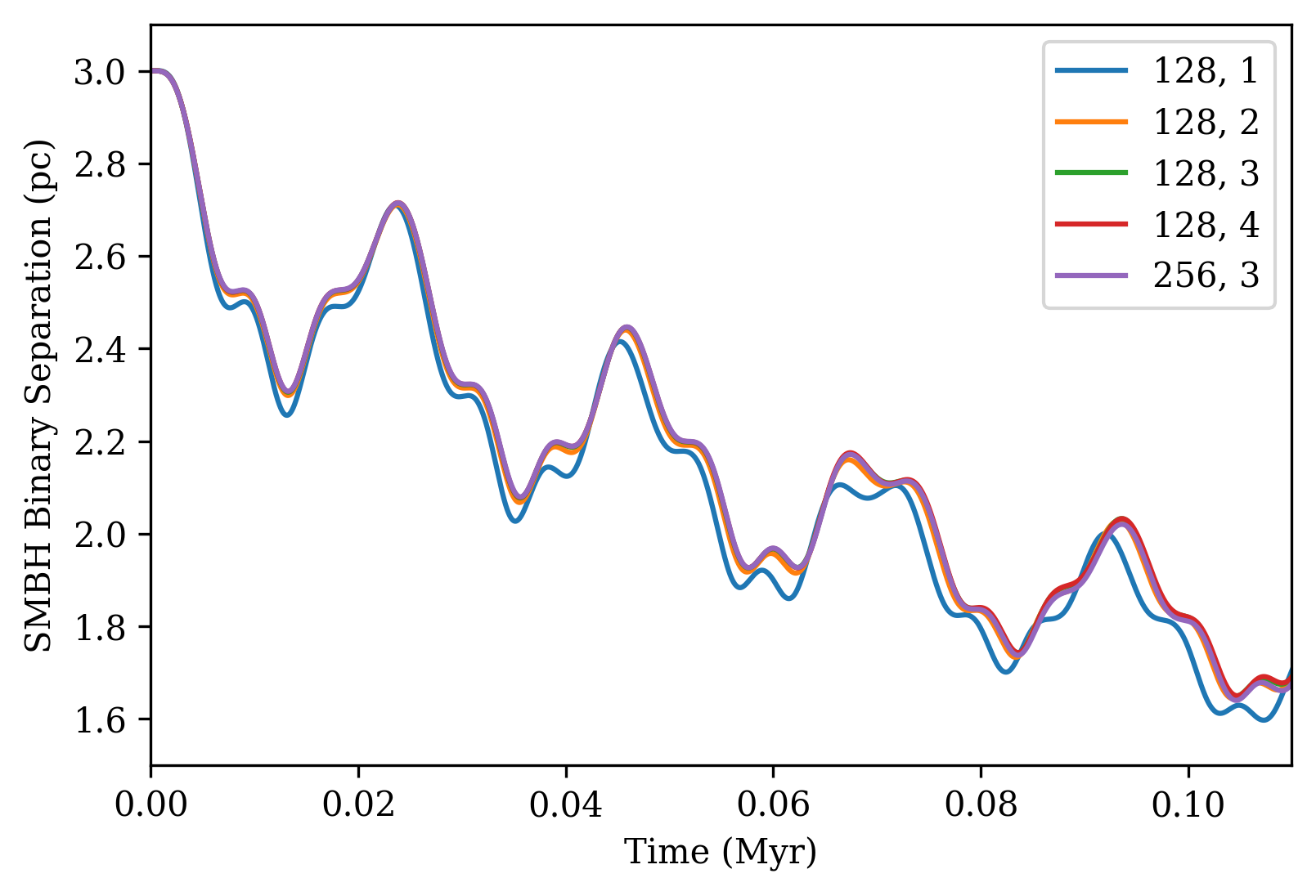}
    \caption{Resolution dependence is illustrated. The label `$128,2$' (for example) indicates a $128^3$ base grid and two levels of refinement. Our default parameters are `$128,3$'.}
    \label{fig:Resolution_Dependence}
\end{figure}

We validate our default parameters by testing for sensitivity to the spatial resolution, timestep, Plummer radius, box size, and the width of each refinement level.\footnote{These tests use $5\times 10^7 \textrm{M}_{\odot}$ black holes;  larger SMBH require better resolution for  energy conservation, as discussed below.} We vary each of these parameters separately while holding the others fixed.  Higher spatial resolution improves accuracy, but must be balanced against computational cost. Figure \ref{fig:Resolution_Dependence} shows that results obtained using our default gridding  match those obtained using (a) a fourth level of refinement or (b) three levels of refinement on a $256^3$ base grid, both of which result in maximum resolution of $2048^3$. A $128^3$ base grid with only two levels of refinement is also largely convergent with the higher resolution runs. However, the lowest resolution run with a single refinement level deviates noticeably from the other schemes. We are therefore satisfied that our default spatial resolution settings are sufficient.

\begin{figure}[tb]
    \centering
    \includegraphics[width=\linewidth]{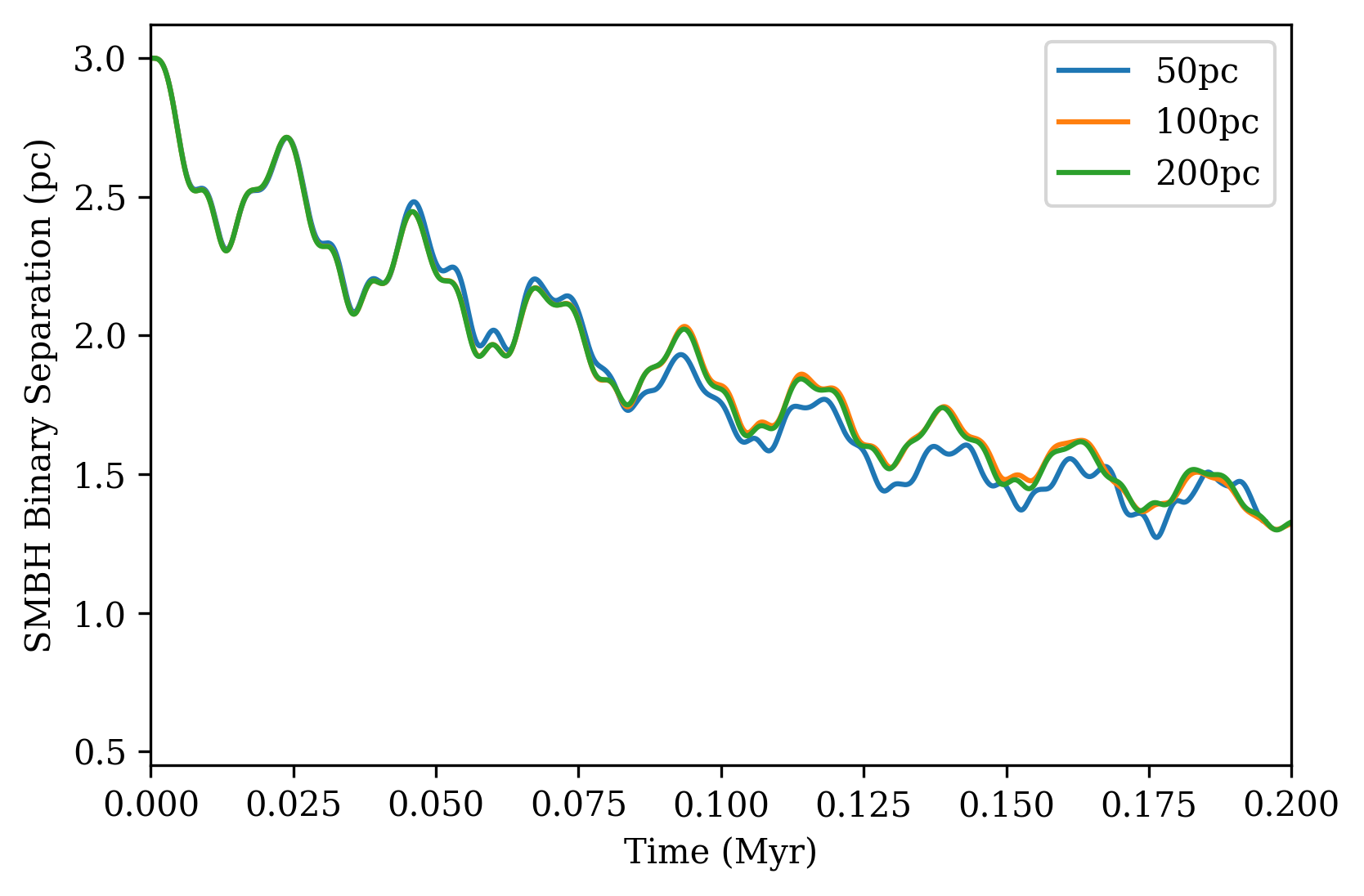}
    \caption{Dependence on the size of the box. }
    \label{fig:Boxsize_Dependence}
\end{figure}

\begin{figure}[tb]
    \centering
    \includegraphics[width=\linewidth]{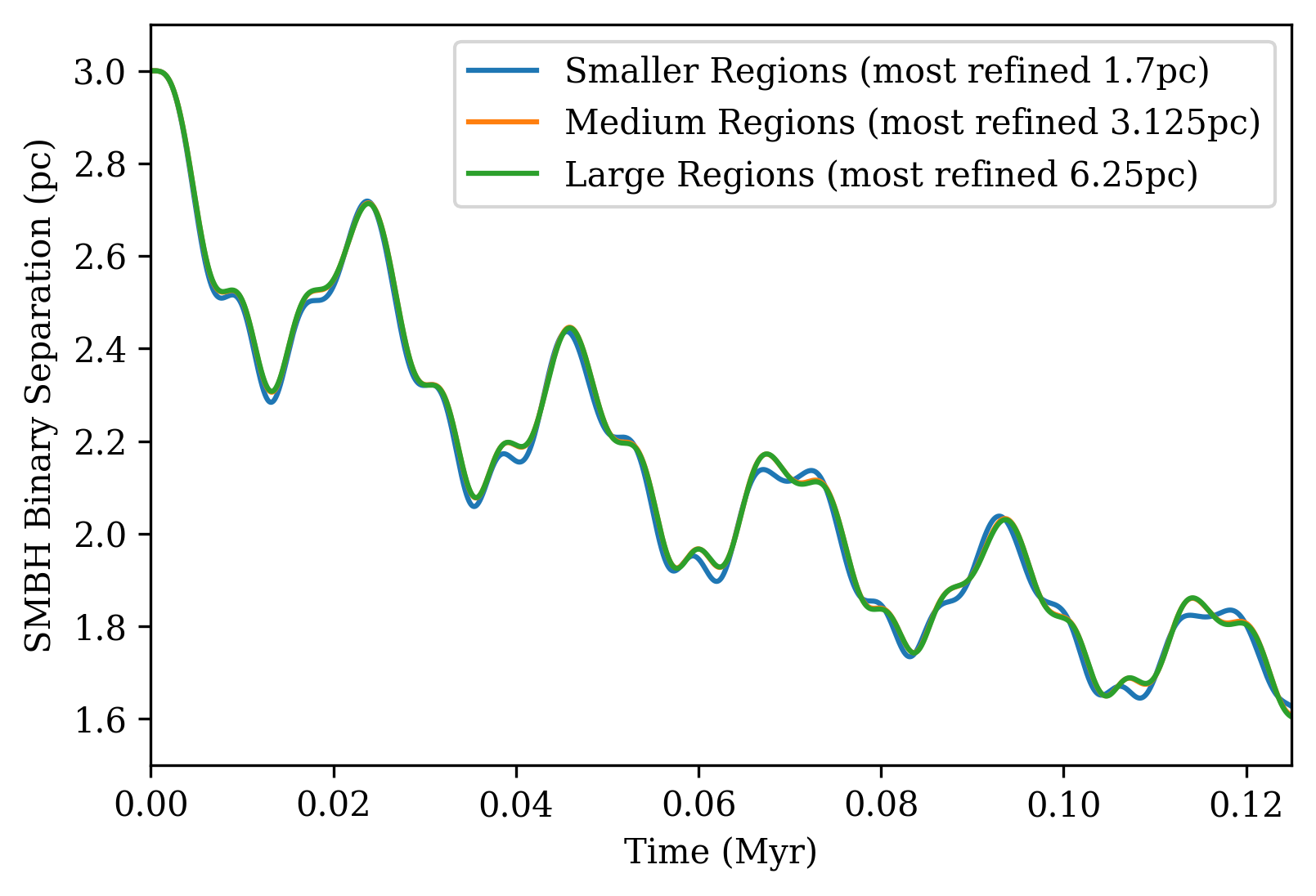}
    \caption{Dependence on the initial size of the maximally resolved ($1024^3$) refinement region, indicated here by the distance of the refinement boundary from the center of the grid. }
    \label{fig:Refined_Region_Dependence}
\end{figure}

\begin{figure}[tb]
    \centering
    \includegraphics[width=\linewidth]{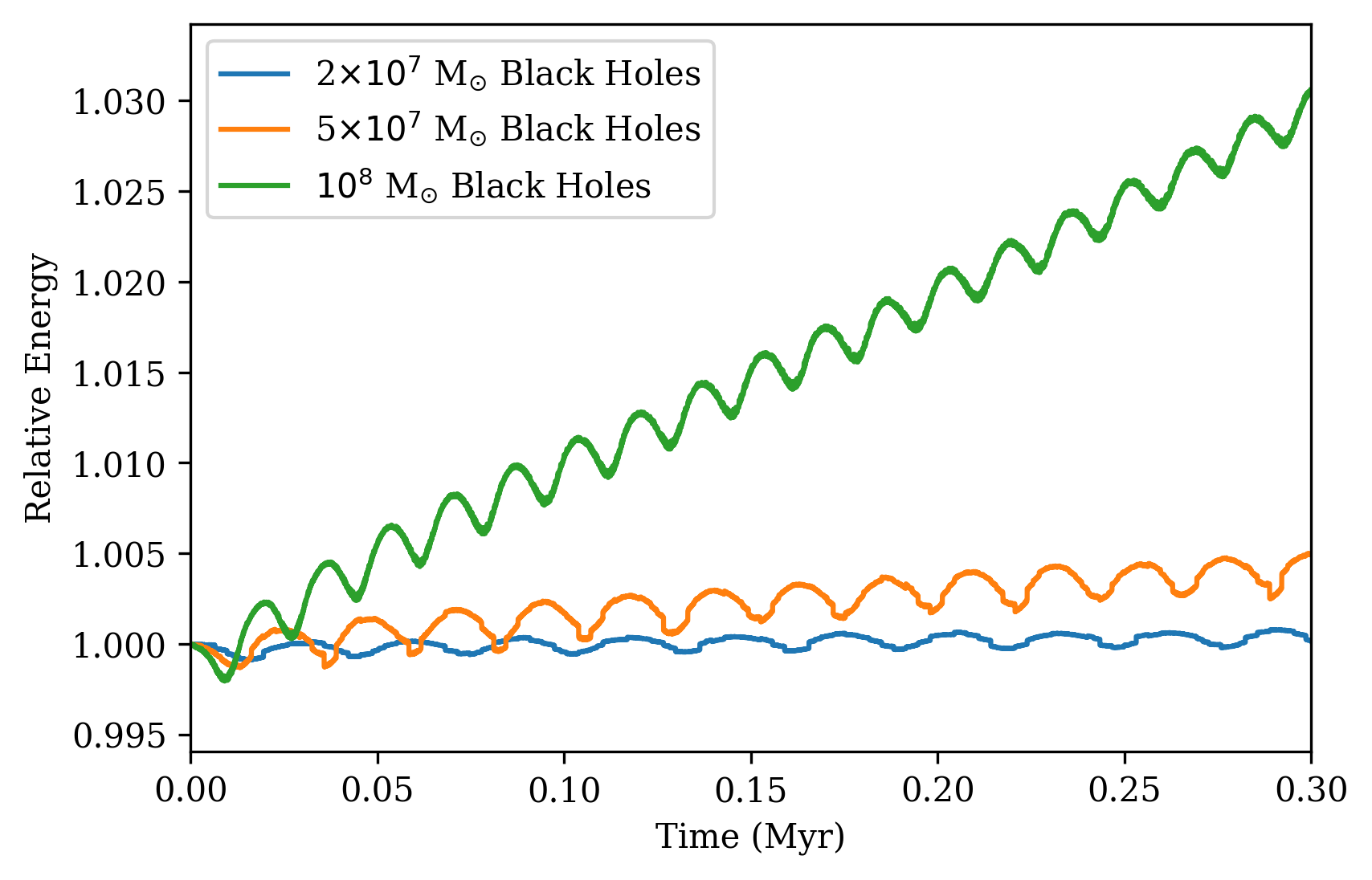}
    \caption{Energy conservation for different mass black holes. Energy conservation worsens with increasing mass, but is still within a range of 1.5\%  over the range of our simulations, while it is significantly worse in the case of the $10^8 \textrm{M}_{\odot}$ black hole.}
    \label{fig:Mass_Energy_Con}
\end{figure}

\begin{figure}[tb]
    \centering
    \includegraphics[width=\linewidth]{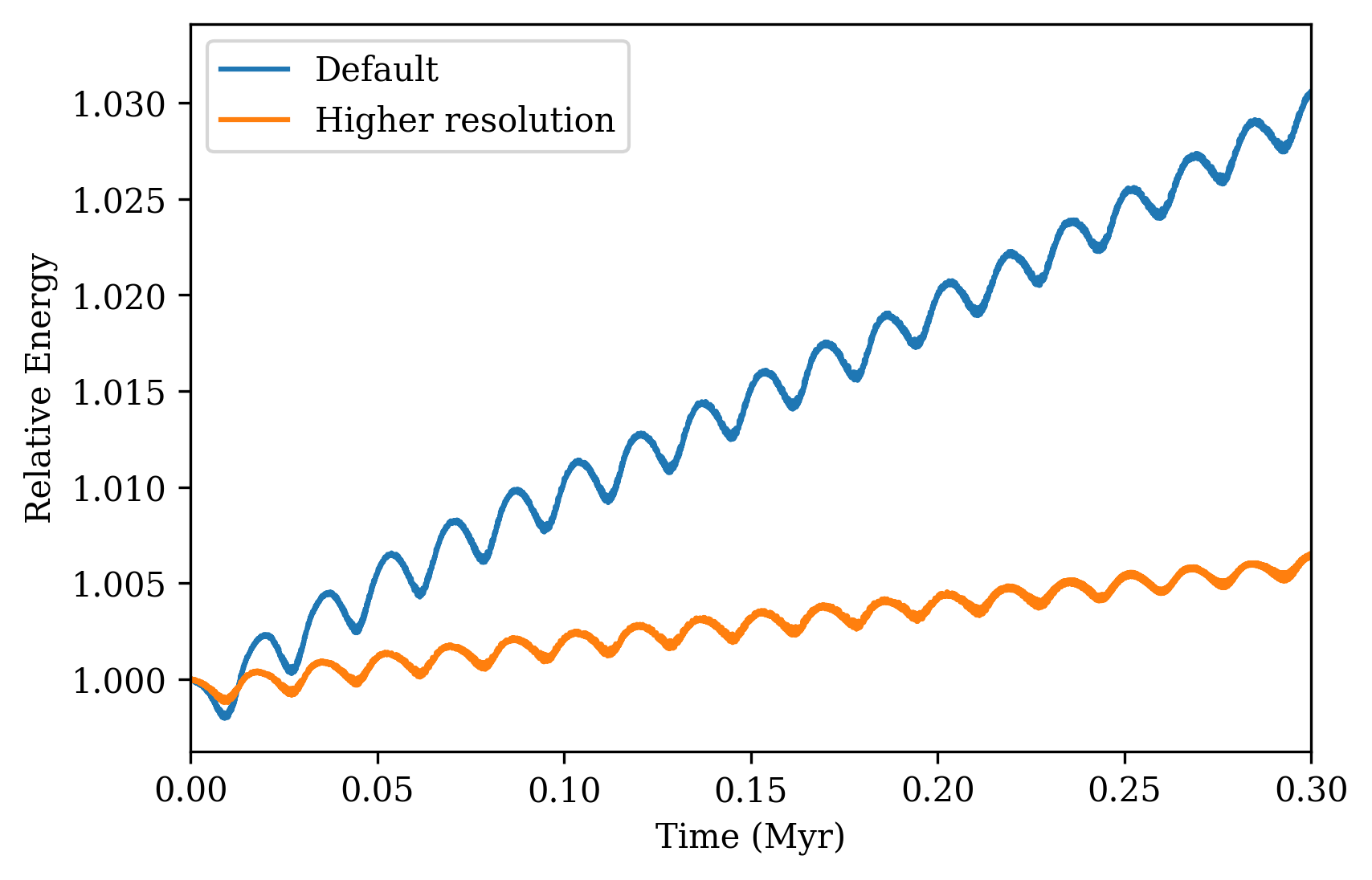}\\
     \includegraphics[width=\linewidth]{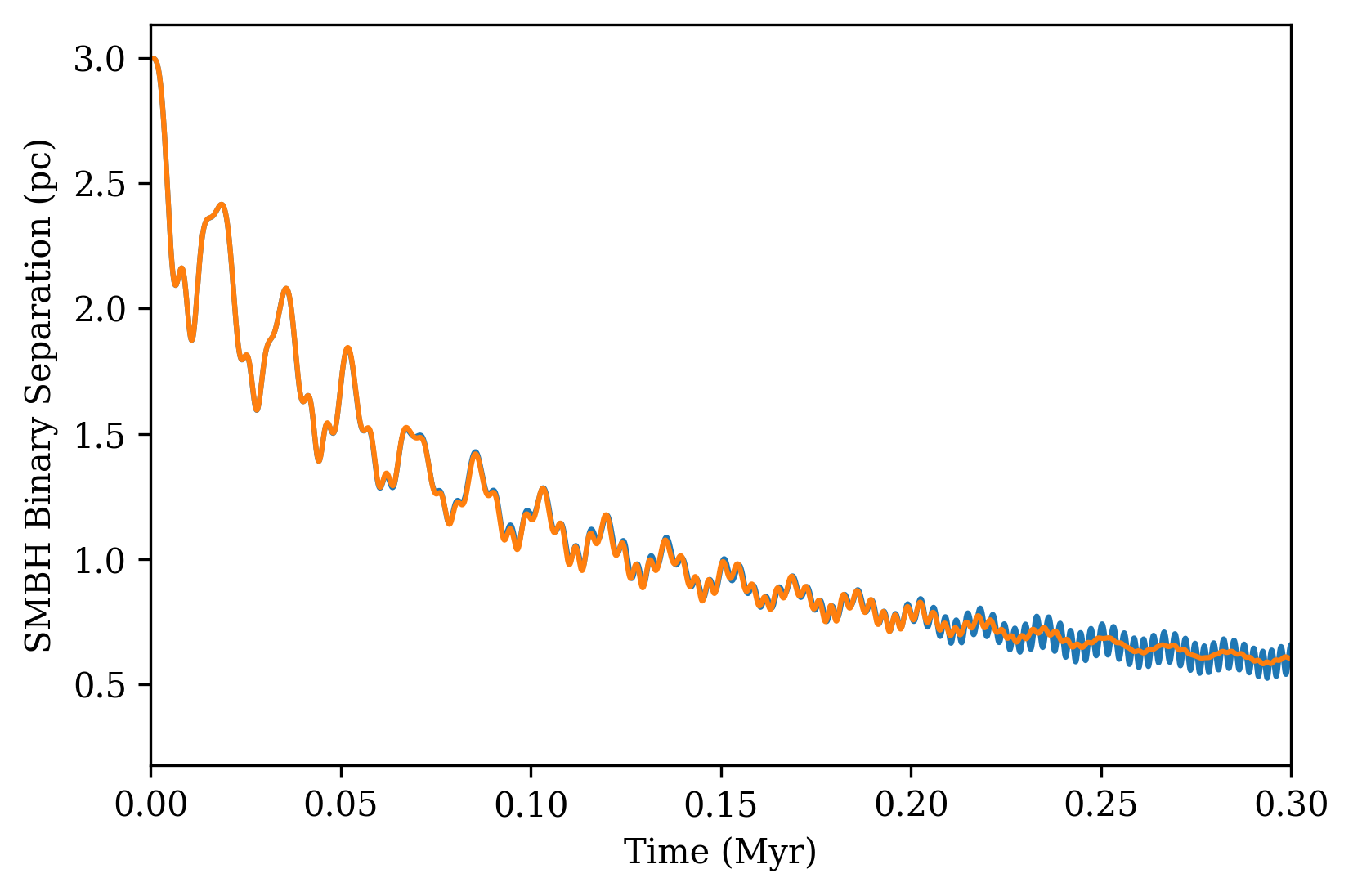}
    \caption{Global energy conservation (top) and separation (bottom) for the $10^8\mathrm{M}_\odot$ binary. Blue curve: $128^3$ base grid with refinement boundaries at (12.5, 6.25, 3.125)pc. Orange curve: $256^3$ base grid with refinement boundaries at (25, 3.125)pc.}
    \label{fig:energy_conservation}
\end{figure}

We carried out a limited run (due to the computational cost) with a timestep parameter $ \sigma^{\textrm{CFL,hyp}} = 0.12$  (or 0.15 times the default value) and saw no improvement. Similarly, we set the Plummer radius to both twice and half its default value and saw no notable differences.   We tested box sizes corresponding to half and twice the default size (at constant physical resolution) to verify that the default box size is sufficiently large to suppress the spurious effects of periodic boundary conditions. Figure \ref{fig:Boxsize_Dependence} shows good convergence between the 100pc and 200pc boxes but the 50pc box diverges from the other two. We therefore find the 100pc default sufficient for our simulations.

By design, the SMBH binary resides within the highest refinement level throughout the simulation and we explore the sensitivity of the SMBH binary evolution to the width of this region in Figure \ref{fig:Refined_Region_Dependence}. If the boundary is very close to the initial positions of the black holes (1.7pc from the center of the grid) we see a small disparity in the evolution of the SMBH binary separation but a boundary 3.125pc from the center of the box is sufficient for convergence. We adopt this value throughout our simulations and place successive boundaries at 6.25pc and 12.5pc. We re-grid periodically to keep the density at the boundaries roughly fixed, as this avoids the buildup of numerical instabilities. 

The default parameters  are suitable for simulations in which the black hole masses are less than $\sim5\times10^7\mathrm{M}_{\odot}$, or 5\% of the soliton mass; larger black holes can lead to discrepancies in mass and energy conservation. This is illustrated in Figure \ref{fig:Mass_Energy_Con}, where the $1\times10^8\mathrm{M}_{\odot}$ run exhibits notably worse energy conservation than the lower mass runs.
We attribute this to the distortion of the soliton induced by the larger black hole, which causes a greater flow of mass across refinement boundaries, leading to accumulation of numerical errors in integrated mass and energy density. To ameliorate this either the base resolution or the size of the refinement regions must be increased. This is illustrated in Figure \ref{fig:energy_conservation} for black holes of mass $10^8\mathrm{M}_{\odot}$; there is a marked improvement in energy conservation when the base resolution is increased to $256^3$ and two refinement levels are included rather than three (the resolution of the maximally refined region is unchanged at $1024^3$). In this scheme the outermost refinement boundary is shifted to 25pc from the center of the box, while the innermost boundary is fixed at 3.125pc from the center. We refer to this combination of settings `higher resolution'.  

Further increases to grid resolution, either by increasing the size of refined regions or base resolution, naturally lead to further improvements in energy conservation. However, our main focus in this work is ascertaining the decrease in the average binary separation over time. We find that while the high-frequency periodic modulation in the binary separation is sensitive to resolution settings, the mean behavior is unchanged. This is illustrated the lower panel of Figure \ref{fig:energy_conservation}. To ensure that our results are robust, we use the `higher resolution' scheme ($256^3$ base resolution with 2 refinement levels) for the $10^8 M_{\odot}$ binary specifically. This suffices for the present work but further investigation is required to determine whether the high frequency modulation is the separation is a numerical artifact.

\section{\label{sec:results}Results}

\begin{figure}[tb]
    \centering
    \includegraphics[width=\linewidth]{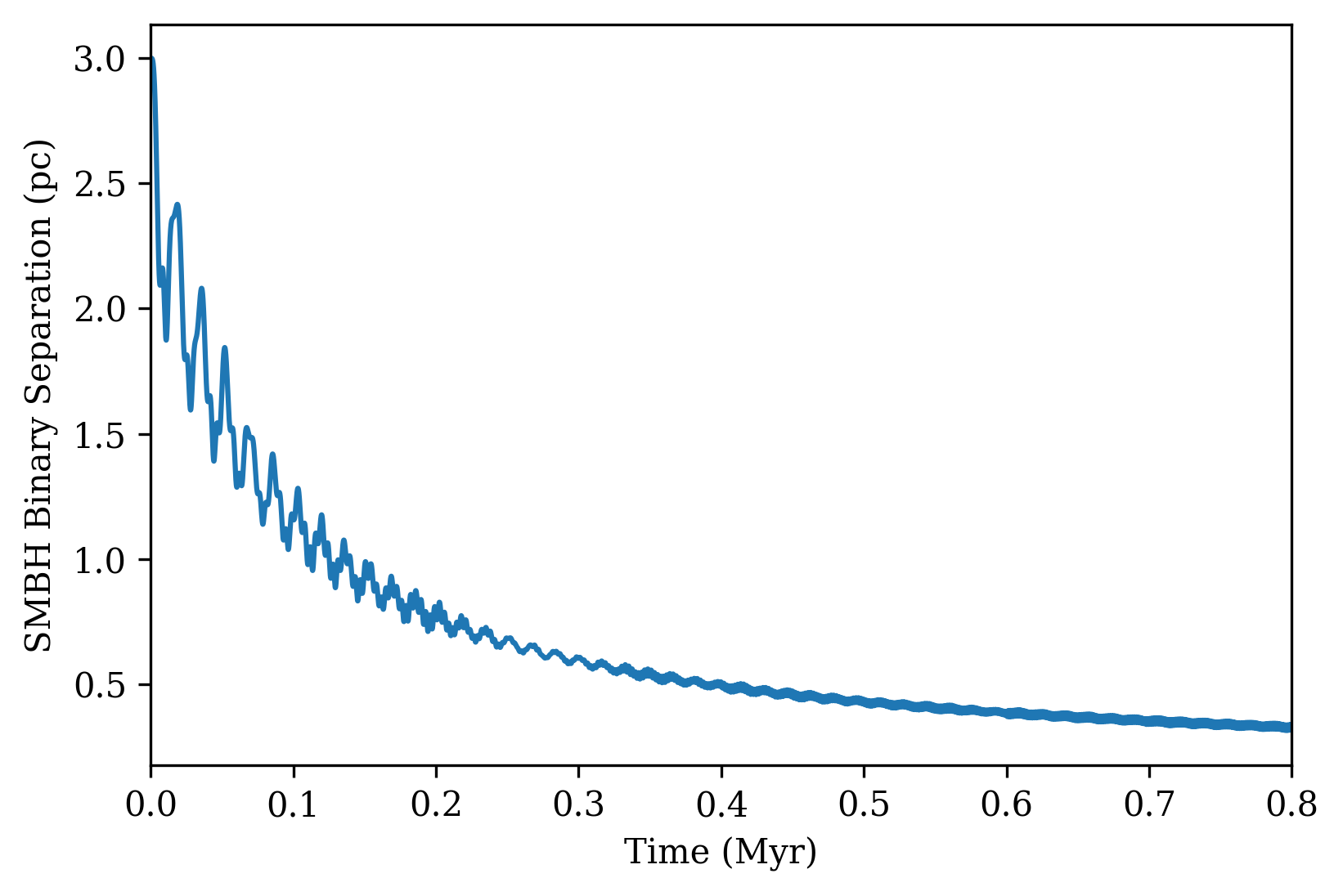}
    \caption{The orbital decay of the SMBH binary using our fiducial simulation parameters.}
    \label{fig:Fiducial_Run}
\end{figure}

\begin{figure}[tb]
    \centering
    \includegraphics[width=\linewidth]{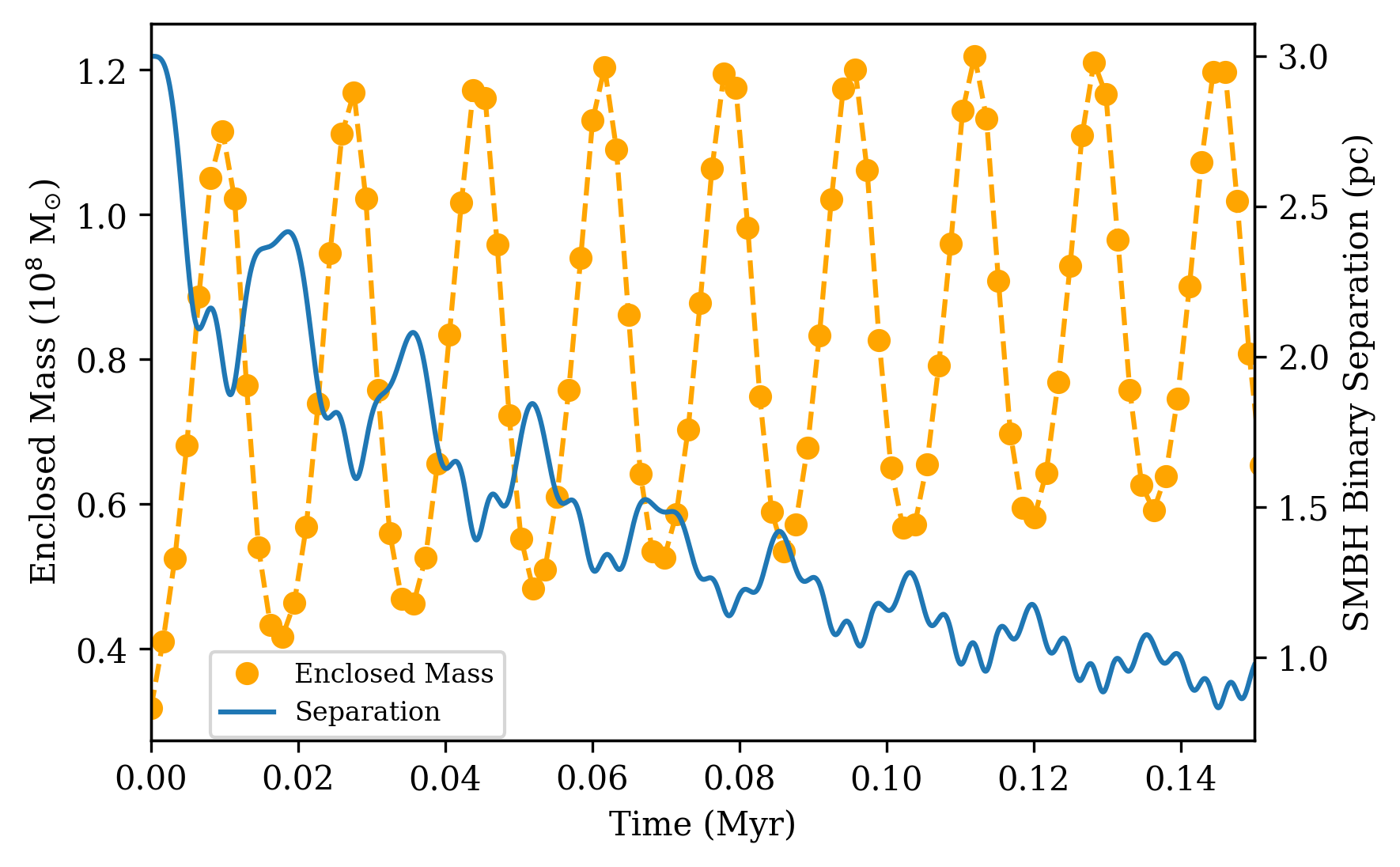}
    \caption{The separation of the black holes plotted with the mass  within 1pc of the soliton center.  The frequency of the  oscillations in SMBH binary separation matches that of  the soliton breathing. (The full box is infrequently written to disk so the breathing plot is interpolated.)}
    \label{fig:Soliton_Breathing}
\end{figure}

Figure \ref{fig:Fiducial_Run} shows the binary separation for the fiducial system over a period of 0.8 Myr. The rate of decay decreases with time but there is a consistent decrease in the orbital separation, bringing the  separation to well below one parsec. The mean separation  is modulated by oscillations induced by a `breathing mode' excited in the soliton. This is illustrated in Figure \ref{fig:Soliton_Breathing}, where we see that the mass contained within the innermost parsec varies periodically. As  the soliton oscillates, the  mass interior to the black holes changes, rendering the velocities of the black holes alternately too high or too low to undertake a circular orbit.

While the oscillation due to the soliton breathing becomes damped as the simulation progresses, a secondary, higher frequency oscillation also appears in the later parts of the simulation. As discussed above this behavior may be in part numerical; the lower panel of Figure \ref{fig:energy_conservation} shows that this oscillation is less noticeable in simulations with better energy conservation. However because this oscillation is subdominant and does not appear to affect the mean separation we do not attempt to fully remove it. 

\begin{figure}
    \centering
    \includegraphics[width=\linewidth]{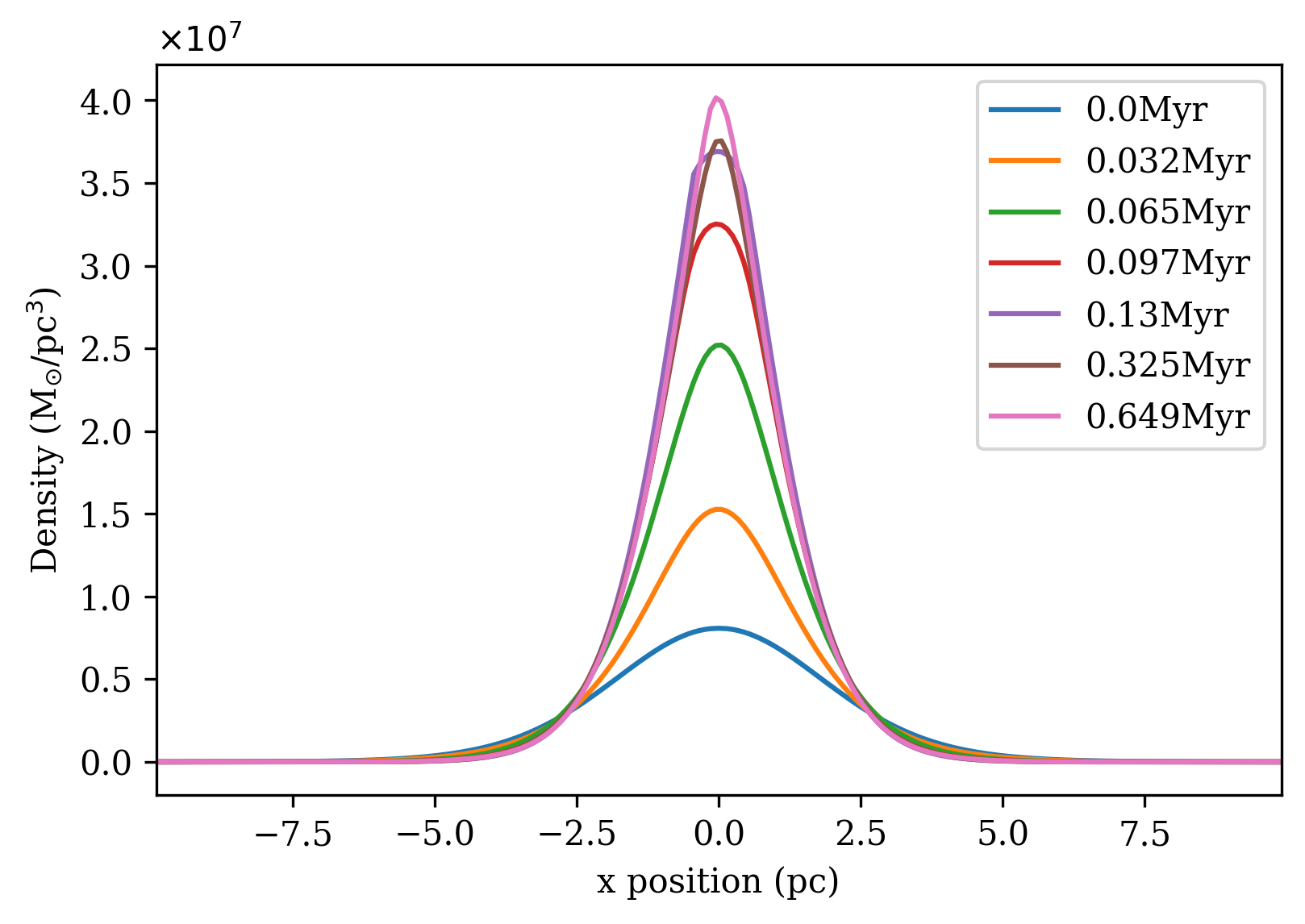}
    \caption{The `pinching' of the ULDM profile during the fiducial run.}
    \label{fig:profile_evolution}
\end{figure}

\begin{figure}[tb]
    \centering
    \includegraphics[width=0.8\linewidth]{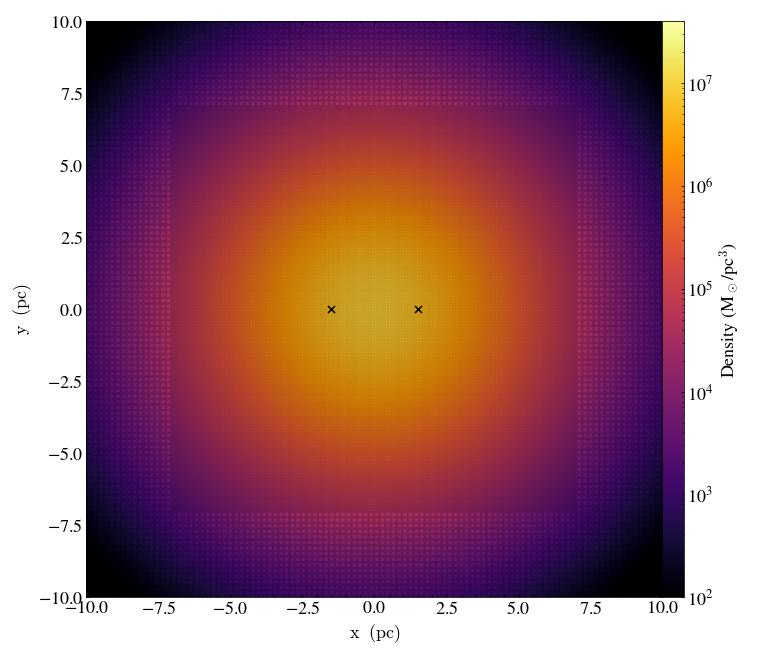}
    \includegraphics[width=0.8\linewidth]{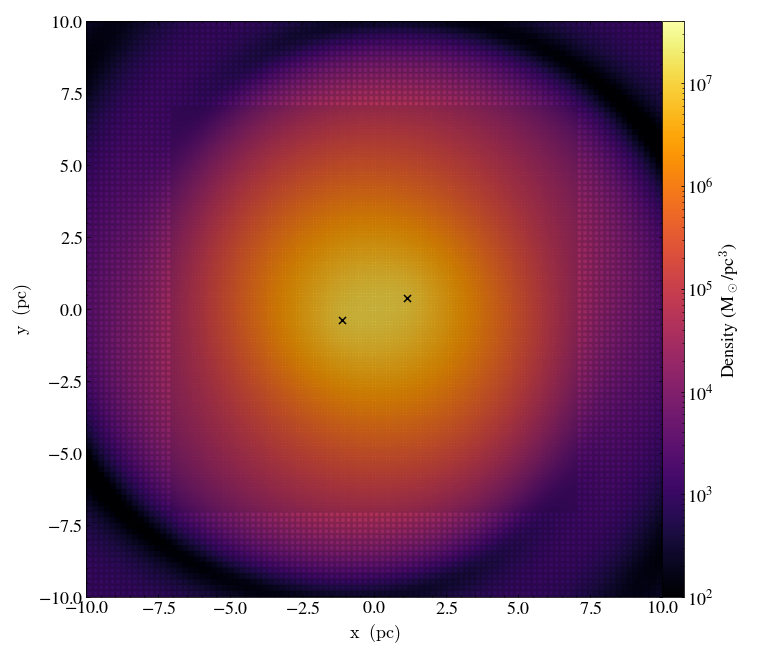}
    \includegraphics[width=0.8\linewidth]{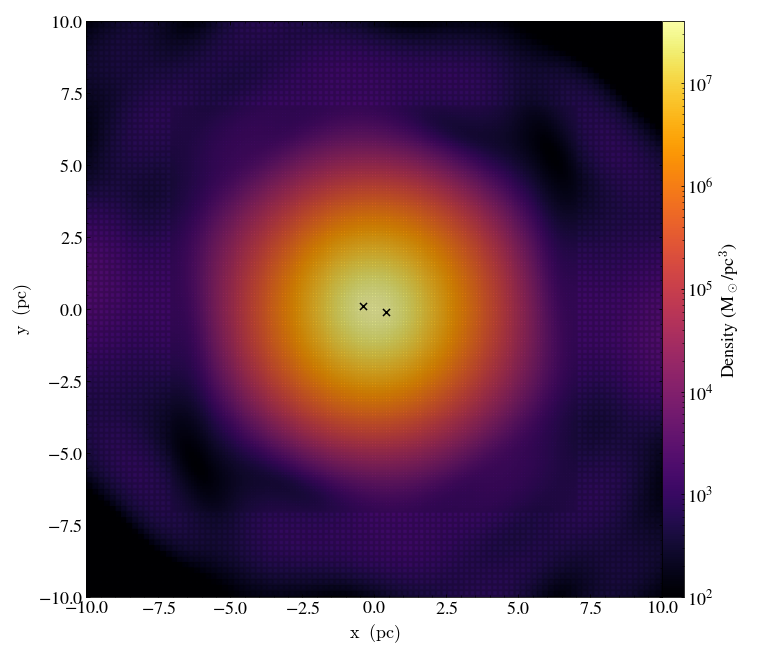}
    \caption{Plots showing the `pinching' of the soliton, at times 0, 0.016, and 0.162 Myr.}
    \label{fig:2d_pinching}
\end{figure}

A further effect not accounted for by simple semi-analytic models is the reshaping of the soliton as the binary approaches the center -- the ULDM density profile becomes increasingly 'pinched' over the course of the simulation. This is due to the gravitational fields of the SMBH, which create a perturbation on the soliton that is not accounted in the initial stable profile. At late times in our fiducial run, the soliton oscillates around a central density of $\sim 4 \times 10^{7} \textrm{M}_{\odot}\textrm{pc}^{-3}$. This is five times higher than the initial central density, while the half-maximum radius falls to $\sim 1.1$pc, half the initial value. This is shown in Figures \ref{fig:profile_evolution} and \ref{fig:2d_pinching}, and will increase the dynamical friction.

\begin{figure}[tb]
    \centering
    \includegraphics[width=\linewidth]{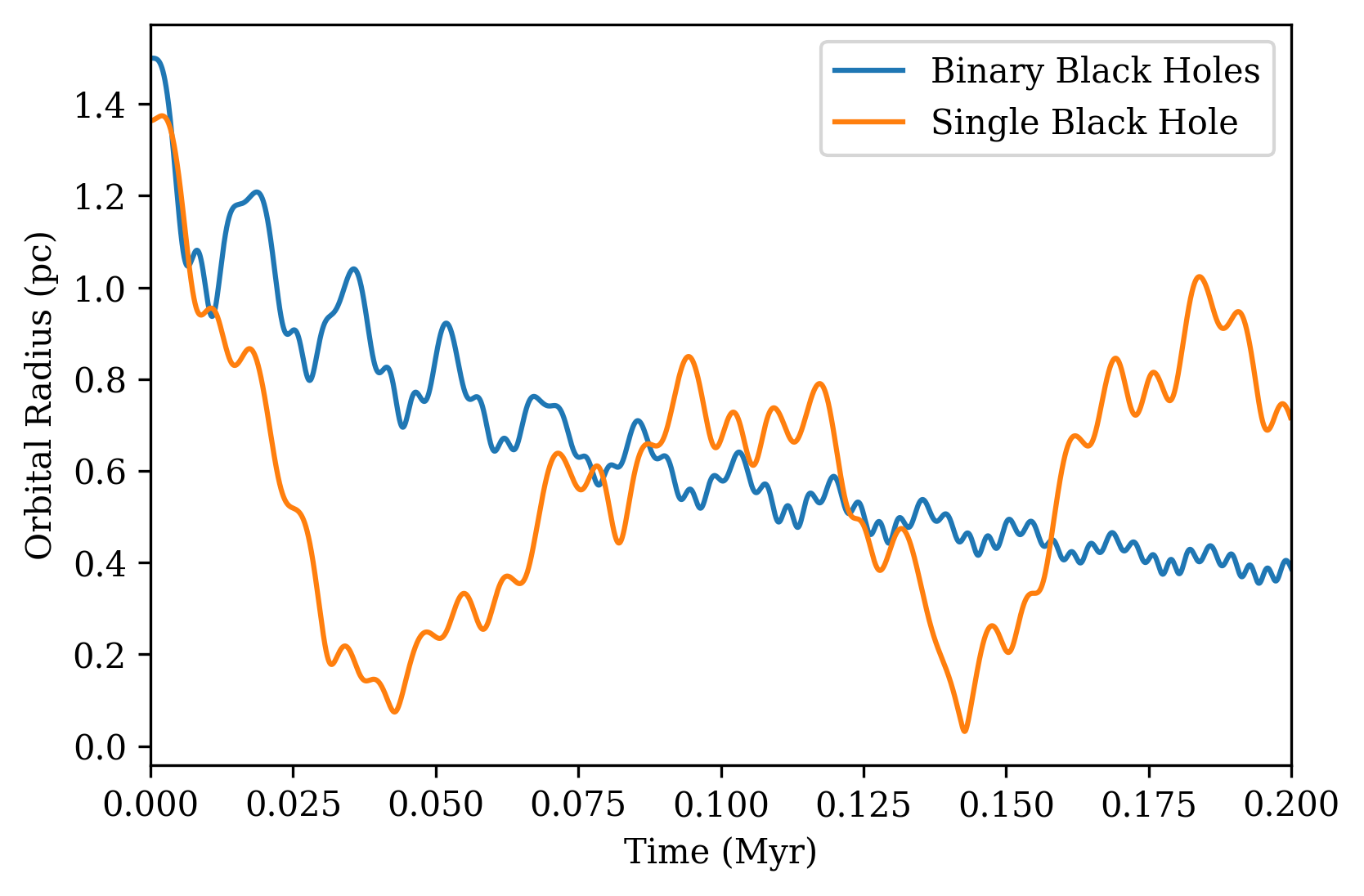}
    \caption{The separation between the SMBH and the center of mass for the single and binary scenarios. The `stone skipping' \cite{Wang:2021udl} is clearly visible in the former but not in the latter.}
    \label{fig:binary_v_single}
\end{figure}

For comparison, Figure \ref{fig:binary_v_single} compares the evolving orbital radius of a single $10^8 \textrm{M}_{\odot}$ black hole to our fiducial binary simulation.\footnote{The soliton center is at the origin so the orbital radius is somewhat smaller. We give the soliton an initial reflex velocity to keep the overall system stationary.} The single black hole undergoes `stone skipping'  \cite{Wang:2021udl}, stalling the migration of the black hole to the center of the soliton. This is not seen in binary systems, for which oscillations are subdominant.\footnote{The difference  presumably arises from the more symmetric binary system coupling to a subset of the modes in the soliton that are excited by a single black hole.}
 
\begin{figure}
    \centering
    \includegraphics[width=\linewidth]{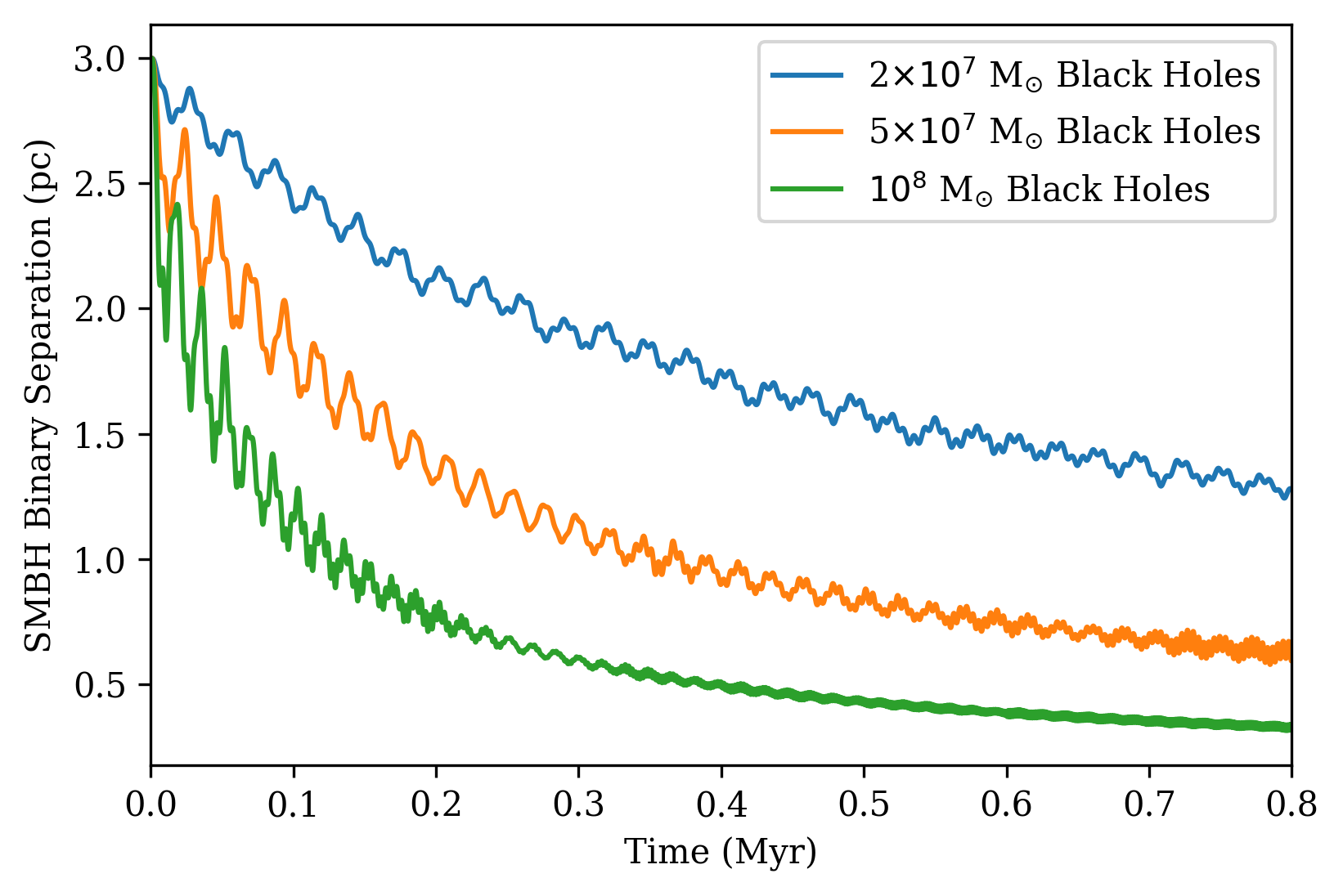}
    \caption{Evolution of the trajectories for different black hole masses.}
    \label{fig:Mass_Dependence}
\end{figure}

In simple models of dynamical friction  \cite{Hui:2016ltb} the drag is proportional to $M^2$, while the initial orbital energy scales with $M$ and  equation (\ref{eq:rdotfull}) indeed scales directly with $M$ whereas the velocity changes are smaller, so we expect larger black holes to decay more quickly. This is confirmed by Figure \ref{fig:Mass_Dependence}, which shows results for black holes with masses from $2 \times 10^7 \textrm{M}_{\odot}$ to $1 \times 10^8 \textrm{M}_{\odot}$  and the ULDM particle and soliton masses with their fiducial values \footnote{Note that the core-halo relation (the relative mass of the soliton to the overall halo) is itself a function of the ULDM particle mass  \cite{chen2024galaxytomographygravitationalwave, Girelli_2020, Schive:2014hza} so this change also implicitly varies the halo mass. } This implies that dynamical friction from ULDM is only relevant to the final parsec problem for very massive black holes, as further discussed in Section \ref{sec:parsec}.

\begin{figure}[tb]
    \centering
    \includegraphics[width=\linewidth]{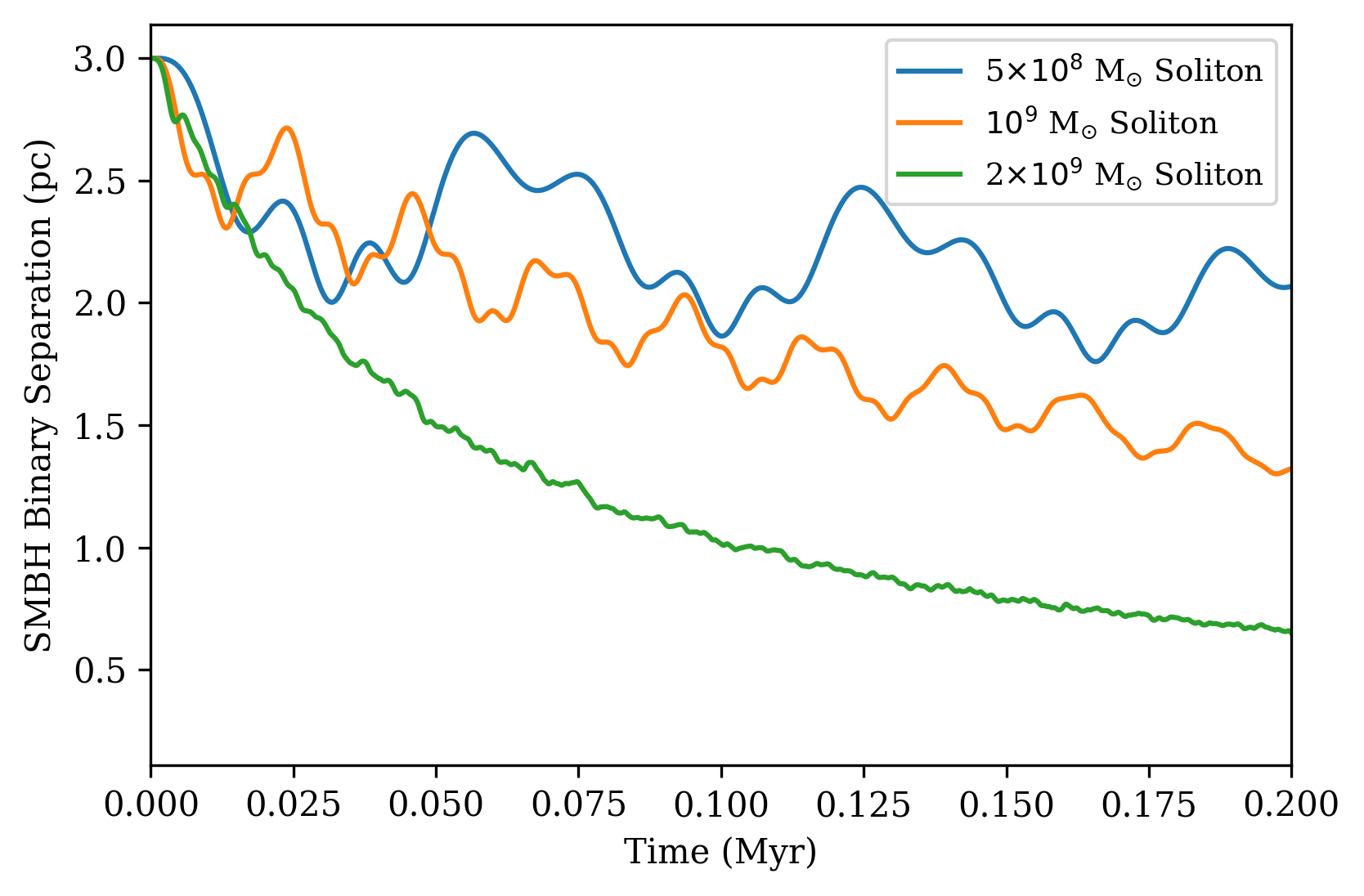}
    \caption{Evolution of the trajectories for half and doubled soliton masses.}
    \label{fig:Soliton_Mass_Dependence}
\end{figure}

Increasing the soliton mass decreases its  size, as the core radius scales as $r_c\propto M^{-1}$. This boosts both the numerator and denominator of equation (\ref{eq:rdotfull}), but the increased density is the dominant effect.  We simulate soliton masses of $5 \times 10^8 \textrm{M}_{\odot}$, $10^9 \textrm{M}_{\odot}$ and $2 \times 10^9 \textrm{M}_{\odot}$, with $5 \times 10^7 \textrm{M}_{\odot}$ black holes and fiducial ULDM particle mass. The results are shown in in Figure \ref{fig:Soliton_Mass_Dependence} and confirm that the decay rate increases with the soliton mass. In addition, the breathing effect is significantly suppressed at larger soliton masses, presumably because radial modes are harder to excite in a more massive soliton. 

\begin{figure}[tb]
    \centering
    \includegraphics[width=\linewidth]{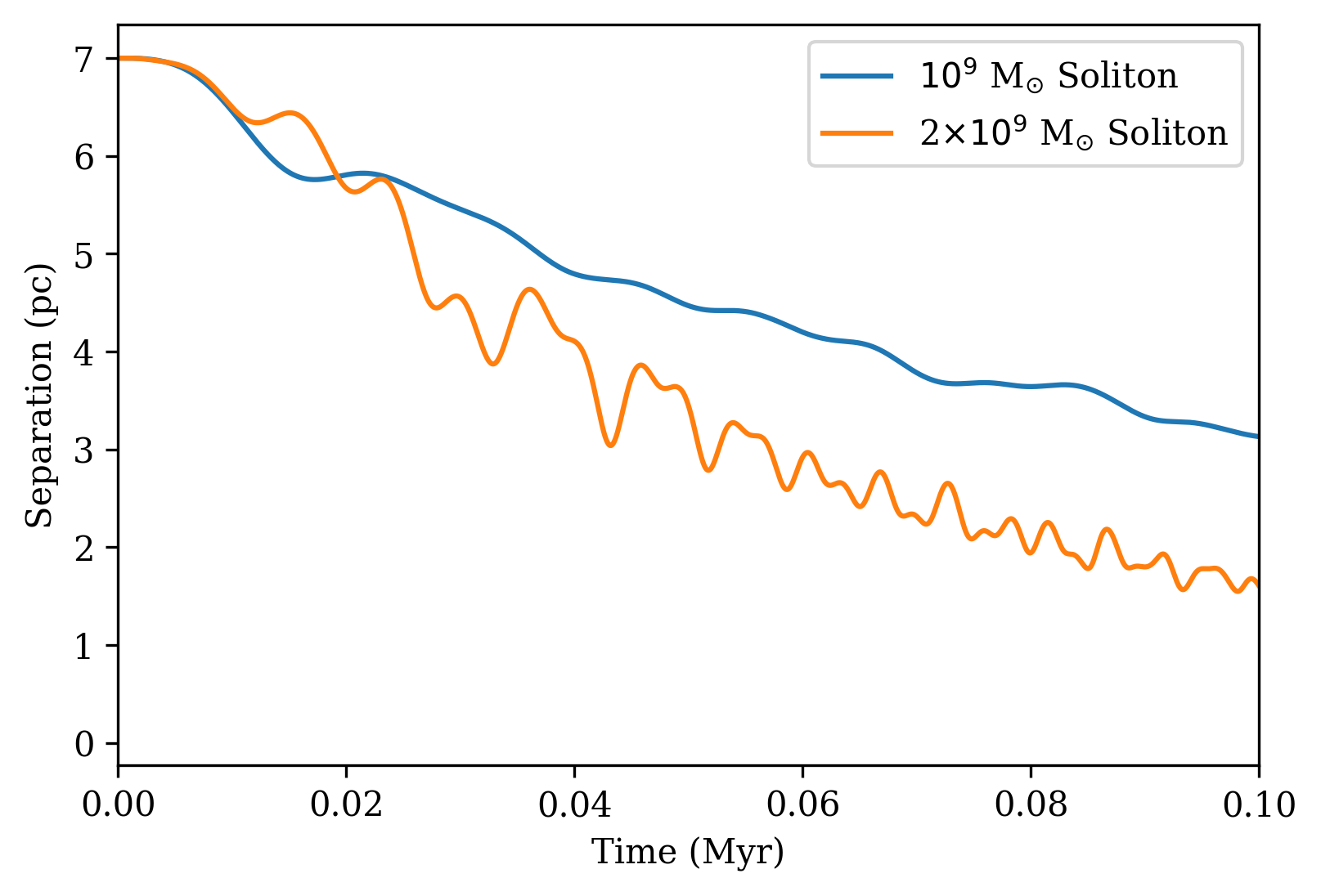}
    \caption{Orbital decay of the black holes from a greater starting distance}
    \label{fig:Soliton_Mass_Dependence_Far}
\end{figure}

Since we include only the soliton our simulations become unrealistic beyond the NFW transition radius, which is 3 or 4 times larger than $r_c$ \cite{Mocz:2017wlg, Kendall:2019fep,Zagorac:2022xic}.  Interestingly, Figure \ref{fig:Soliton_Mass_Dependence_Far} shows that a black hole starting close to the transition radius of a $2\times 10^9 \textrm{M}_{\odot}$ soliton still decays more rapidly that in a $10^9 \textrm{M}_{\odot}$ even though the initial local ULDM density is slightly lower for the more massive soliton. That said, the density of the more massive soliton rises more quickly with radius, so the drag will also increase more rapidly, separating the two trajectories.

\begin{figure}[tb]
    \centering
    \includegraphics[width=\linewidth]{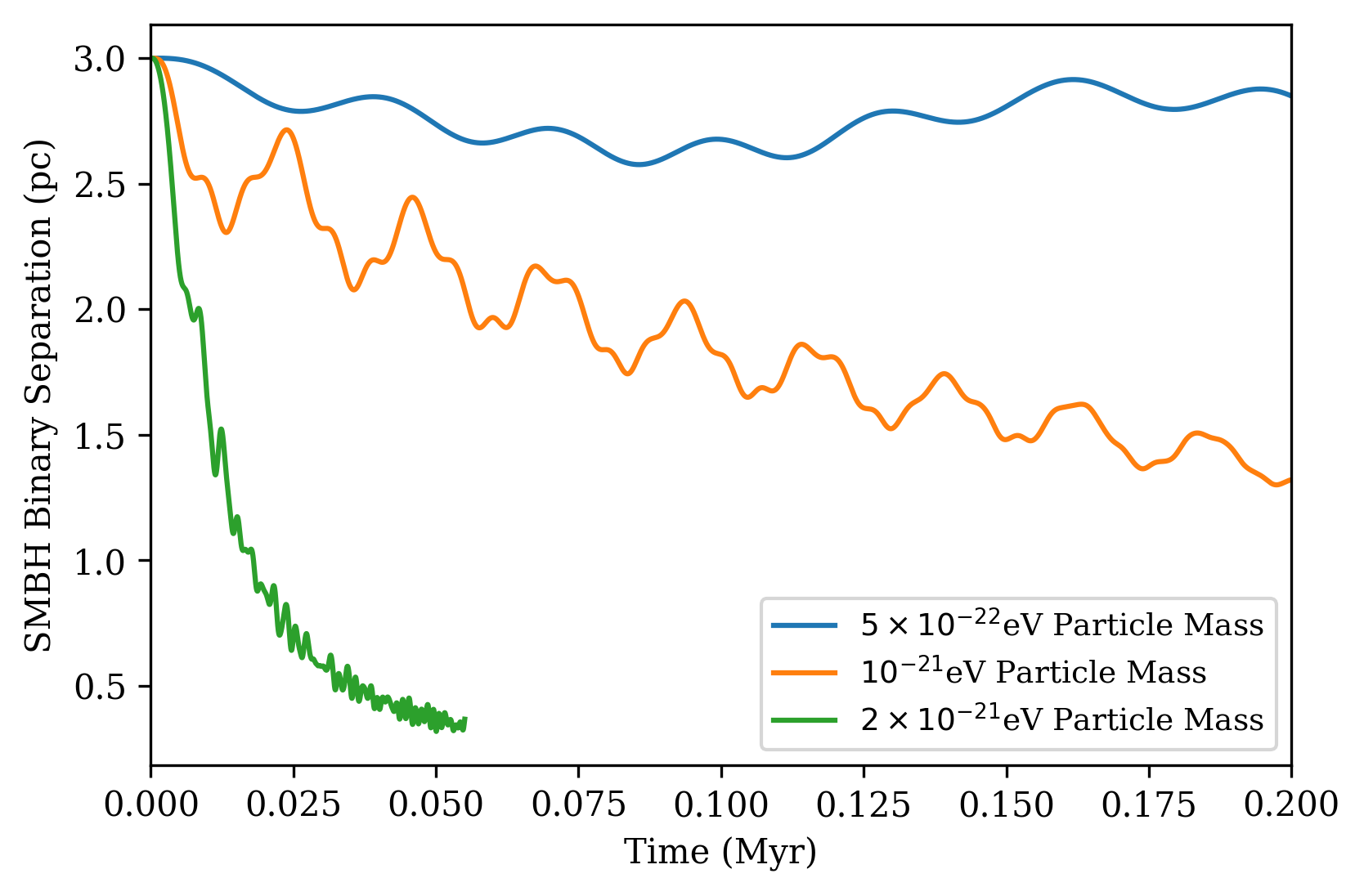}
    \caption{Evolution of the trajectories for half and doubled ULDM particle masses and $5\times 10^7 \mathrm{M}_\odot$ black holes. The $2\times10^{-21}$eV mass simulation is computationally expensive and it is halted after the decay rate has become obvious.}
    \label{fig:ULDM_Particle_Mass_Dependence}
\end{figure}

For a fixed soliton mass, the core radius scales as $r_c\propto m^{-2}$, where $m$ is the ULDM particle mass.  Consequently, the central density  scales as $m^6$. Via equation (\ref{eq:rdotfull}) we see that the dynamical friction is proportional to the density,  so we expect a much more rapid decay at larger values of $m$. This will be offset to some extent by the velocity terms in the denominator but since $v\sim 1/\sqrt{r}$ for a point mass these will not rise as rapidly as the density term in the numerator.  
Figure \ref{fig:ULDM_Particle_Mass_Dependence} compares the evolution of black holes of mass $5\times10^7\textrm{M}_{\odot}$ within a soliton of mass $10^9\textrm{M}_{\odot}$ for different values of $m$ confirming our expectations for decay rate, and we also see that the breathing mode and pinching is less pronounced at larger $m$. The is consistent with the narrower profile and deeper central gravitational potential near the soliton centre at larger $m$, reducing the impact of the black hole potential.

Taken together these results suggest that it is possible for ULDM to significantly modify the decay of an SMBH binary. That said,  any such impact will be most obvious in large solitons and with ULDM particle masses at the upper end of their overall range. 

\section{\label{sec:semi-analytic}Semi-Analytic Fits}

Koo {\em et al.\/} \cite{Koo:2023gfm}  estimate the orbital decay due to the dynamical friction experienced by a binary SMBH inside a soliton. Starting from equation (\ref{eq:Hui_friction}) they find an integrable form for $\dot{r}$ after applying approximations that hold for small separations. As  $M_{enc}$ decreases, Equation (\ref{eq:circmotion}) reduces to  its second term and after the soliton  has  stabilized  (e.g. as shown in Figure \ref{fig:profile_evolution}), the density is roughly constant. Additionally, at small values of $r$, $k\tilde{r}\ll1$ so  the integral for the coefficient of friction may be approximated as $C\approx ({{k\tilde{r}})^2}/{3}$  \cite{Hui:2016ltb,Wang:2021udl}. Equation (\ref{eq:torque})  then becomes
\begin{equation}
    \dot{L}=-Fr=-\frac{4\pi G^2 M^2 m^2 \alpha^2 r^3\rho_0}{3\hbar^2},
    \label{eq:torque-simp}
\end{equation}
where $\rho_0$ denotes the central density.

Recalling that $L=Mvr$ the torque provides  the rate of change in orbital radius,
\begin{equation}
    \dot{L}= \frac{M^{3/2}\sqrt{G}}{4{\sqrt{r}}} \dot{r}.
    \label{eq:dLdt-simp}
\end{equation}
Finally, we can convert this into the rate of change of binary separation, $D$, since $D=2r$, 
\begin{equation}
    \dot{D}=-K D^{7/2},
    \label{eq:drdt-simp}
\end{equation}
where
\begin{equation}
    K=\frac{(2 G)^{3/2} M^{1/2} (m \alpha)^2 \pi\rho_0}{3\hbar^2}.
    \label{eq:K}
\end{equation}
This integrates to give 
\begin{equation}
    D=\left( \frac{5 K}{2} (t-t_0)+{D_0}^{-5/2}\right)^{-2/5},
    \label{eq:drdt-integral}
\end{equation}
where $t_0$ and $r_0$ are our initial time and radius. This matches equation (\ref{eq:rdotfull}) under the condition $v^3\gg 4\pi r^2 \rho G v$, which is true at small $r$.

In deriving Equation (\ref{eq:drdt-integral}), we assume $M_{\textrm{enc}}\ll M$. This  is not well-satisfied at the beginning of the simulation  but  improves quickly as the SMBH sink towards the center of the soliton. Consequently, we fit our fiducial simulation  from 0.162Myr, or 50,000 timesteps into the simulation. At this point, $D<1\textrm{pc}$ and $M_{enc}=1.37 \times 10^7 \textrm{M}_{\odot}$. Accounting for this additional enclosed mass would increase our initial velocity estimate  by $\sim 25\%$.

\begin{figure}[tb]
    \centering
    \includegraphics[width=\linewidth]{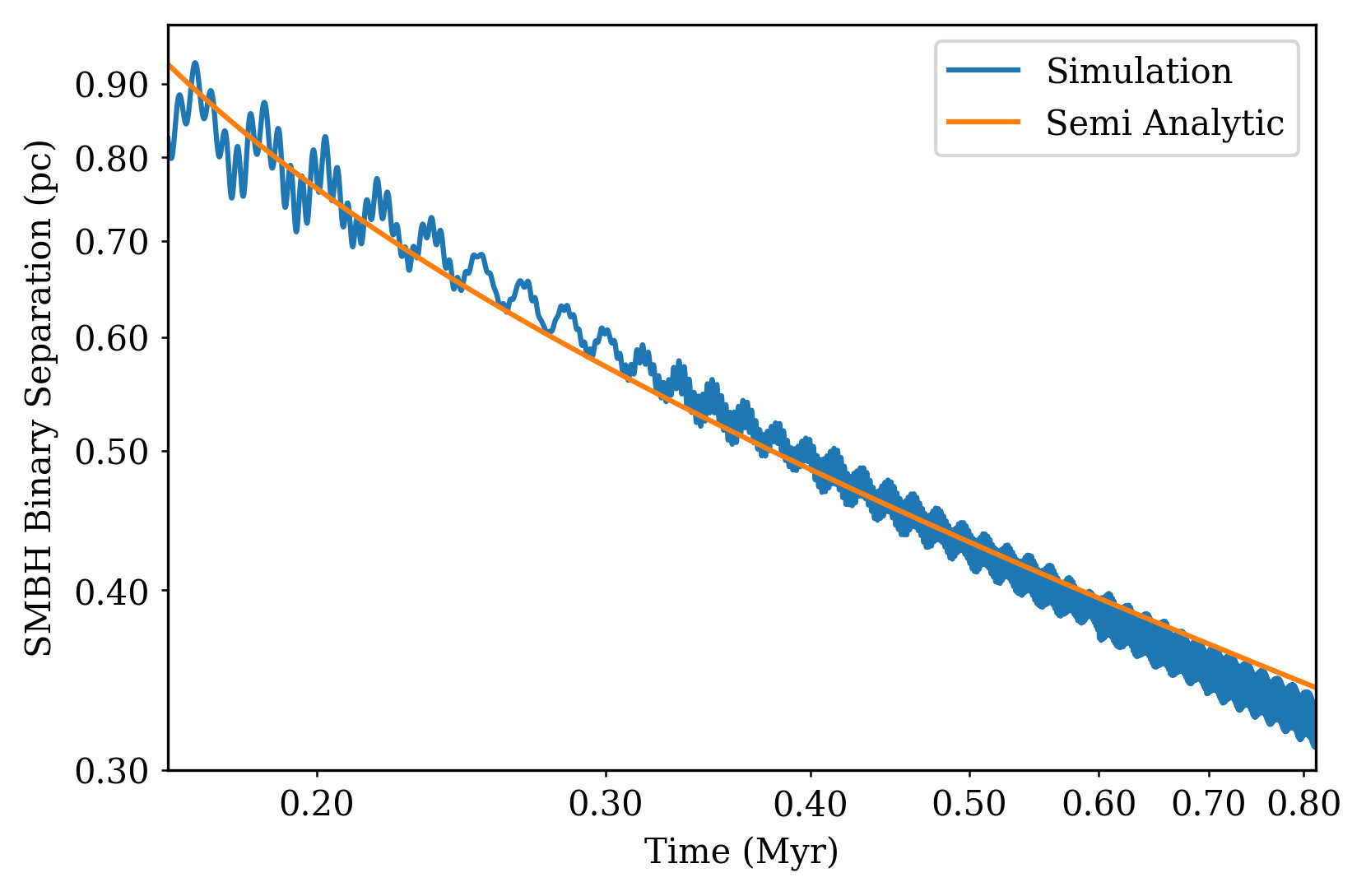}
    
     \includegraphics[width=\linewidth]{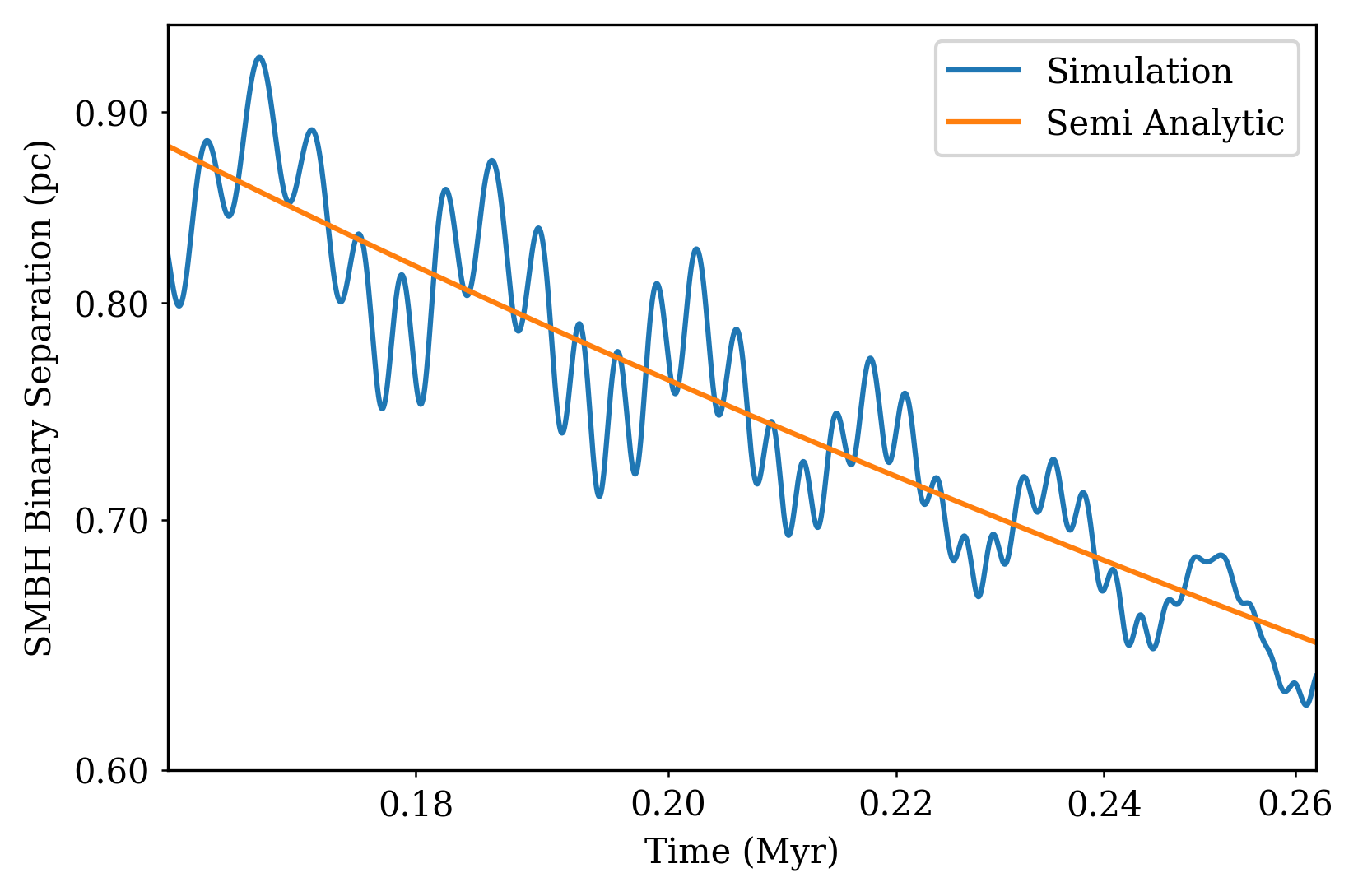}
    \caption{A fit of equation (\ref{eq:drdt-integral}) to our fiducial run (top) and the 100kyr interval from the moment where the separation is  $\sim0.9$pc (bottom).}
    \label{fig:semi-analytic-fit}
\end{figure}

A fit of Equation (\ref{eq:drdt-integral}) to our numerical results is shown in Figure \ref{fig:semi-analytic-fit} and we see that the semi-analytic formula overestimates the orbital decay at early times and underestimates at later times. The free parameter in the fit is 
\begin{multline}
    K=8.22\left(\frac{\rho_0}{3.84\times10^7\textrm{M}_{\odot}\textrm{pc}^{-3}}\right){\left(\frac{M}{10^8\textrm{M}_{\odot}}\right)}^{1/2} \\ \left(\frac{m}{10^{-21}\textrm{eV}}\right)^2\textrm{pc}^{-5/2}\textrm{Myr}^{-1},
    \label{eq:k-val}
\end{multline}
corresponding to $\alpha=0.3027$. This in turn implies an initial value of $k\tilde{r}=0.04$, consistent with $k\tilde{r}\ll1$. Koo {\em et al.\/} \cite{Koo:2023gfm} set $\rho_0$ with reference to the canonical  soliton profile while we use the simulated central density at the beginning of our fit, after the pinching has stabilized. This does not affect the fit to $K$ but will modify the computed value of $\alpha$.

The bottom panel of Figure \ref{fig:semi-analytic-fit} shows only the portion of the evolution analysed by Koo {\em et al.} \cite{Koo:2023gfm}. Since our simulation is already underway when it reaches this radius we do not expect an exact match  but the qualitative features agree well.  The numerical fit appears to be perfectly acceptable in this  limited region, highlighting the value of the longer runs.

\begin{figure}[tb]
    \centering
    \includegraphics[width=\linewidth]{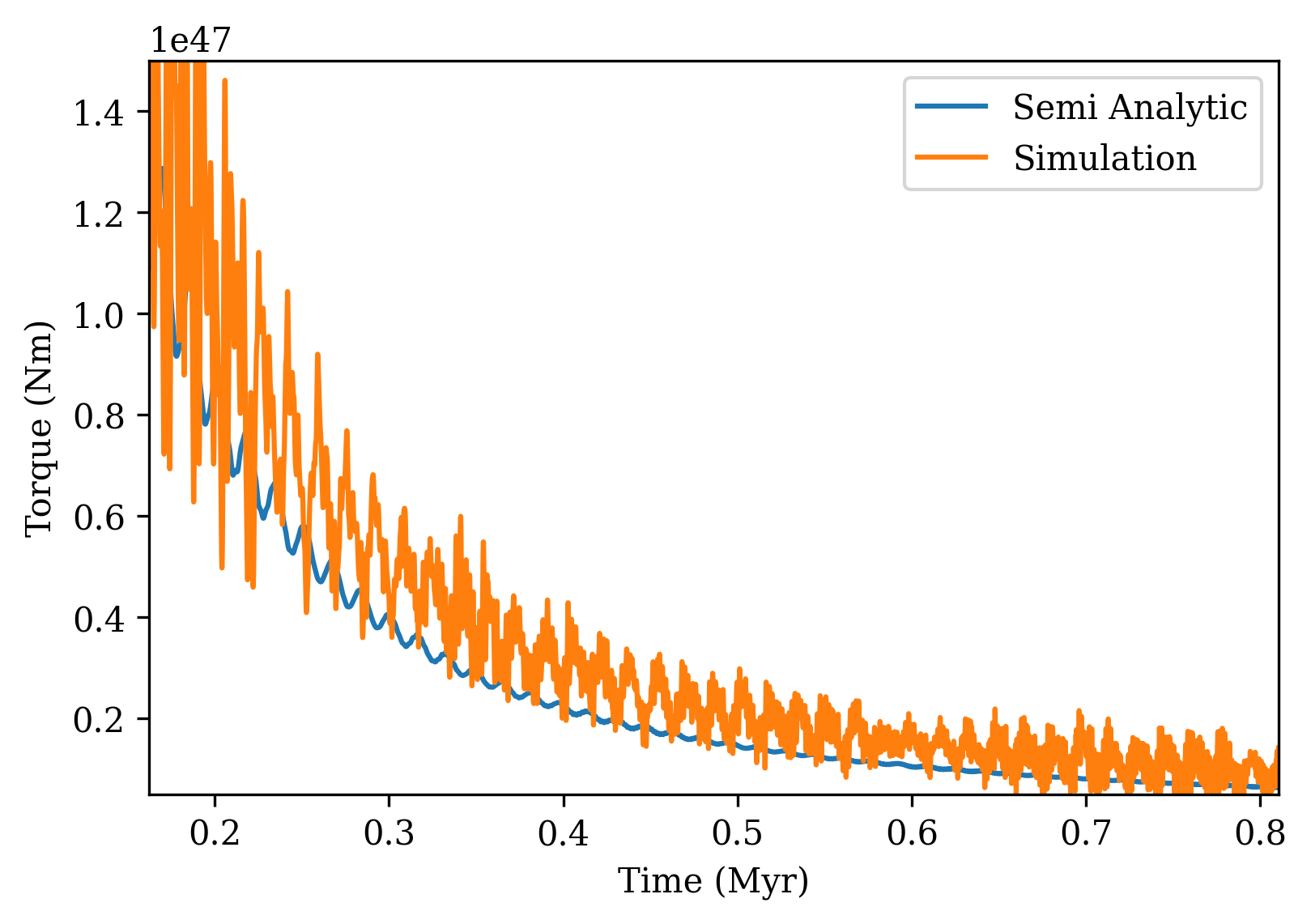}
    \includegraphics[width=\linewidth]{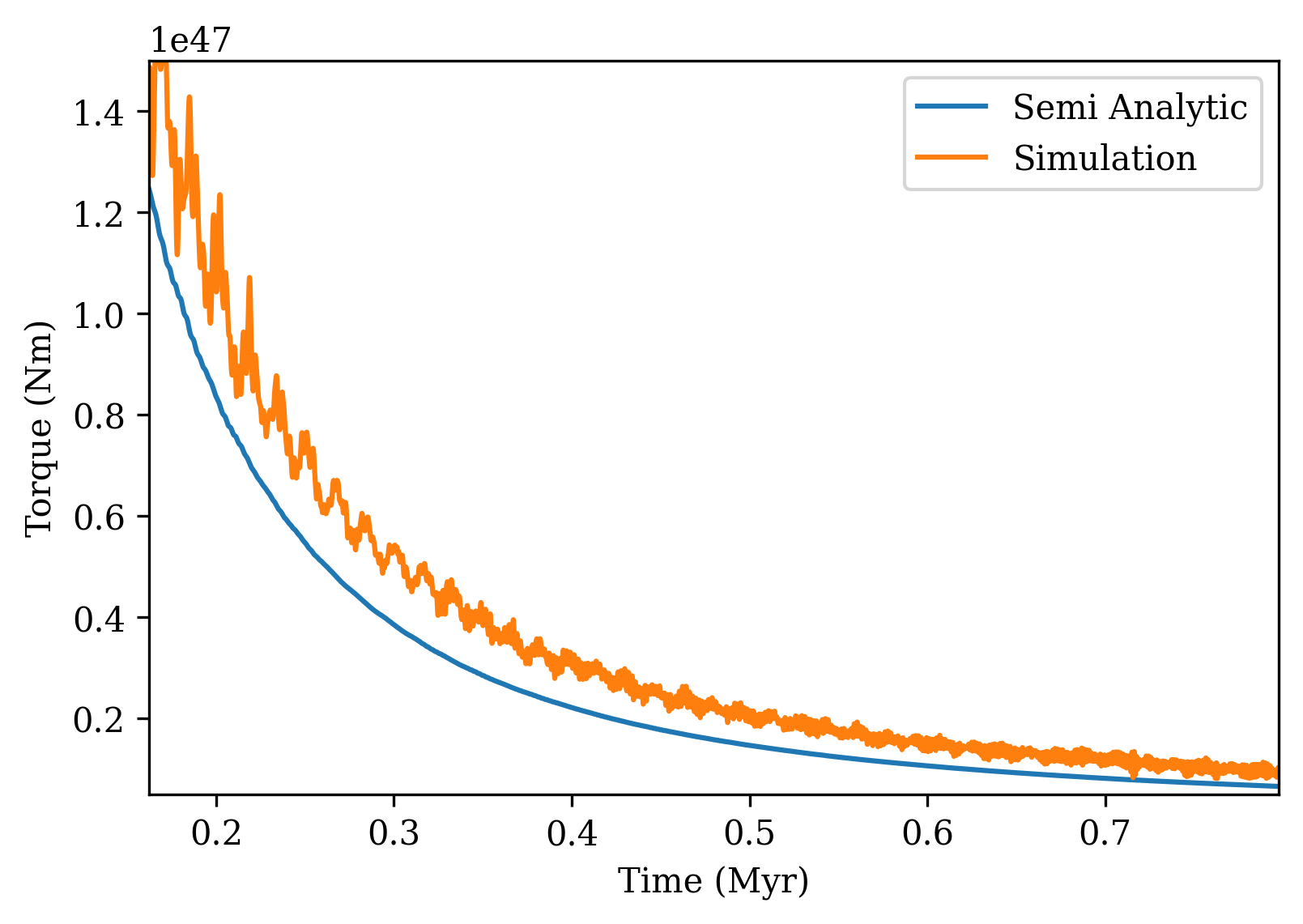}
    \includegraphics[width=\linewidth]{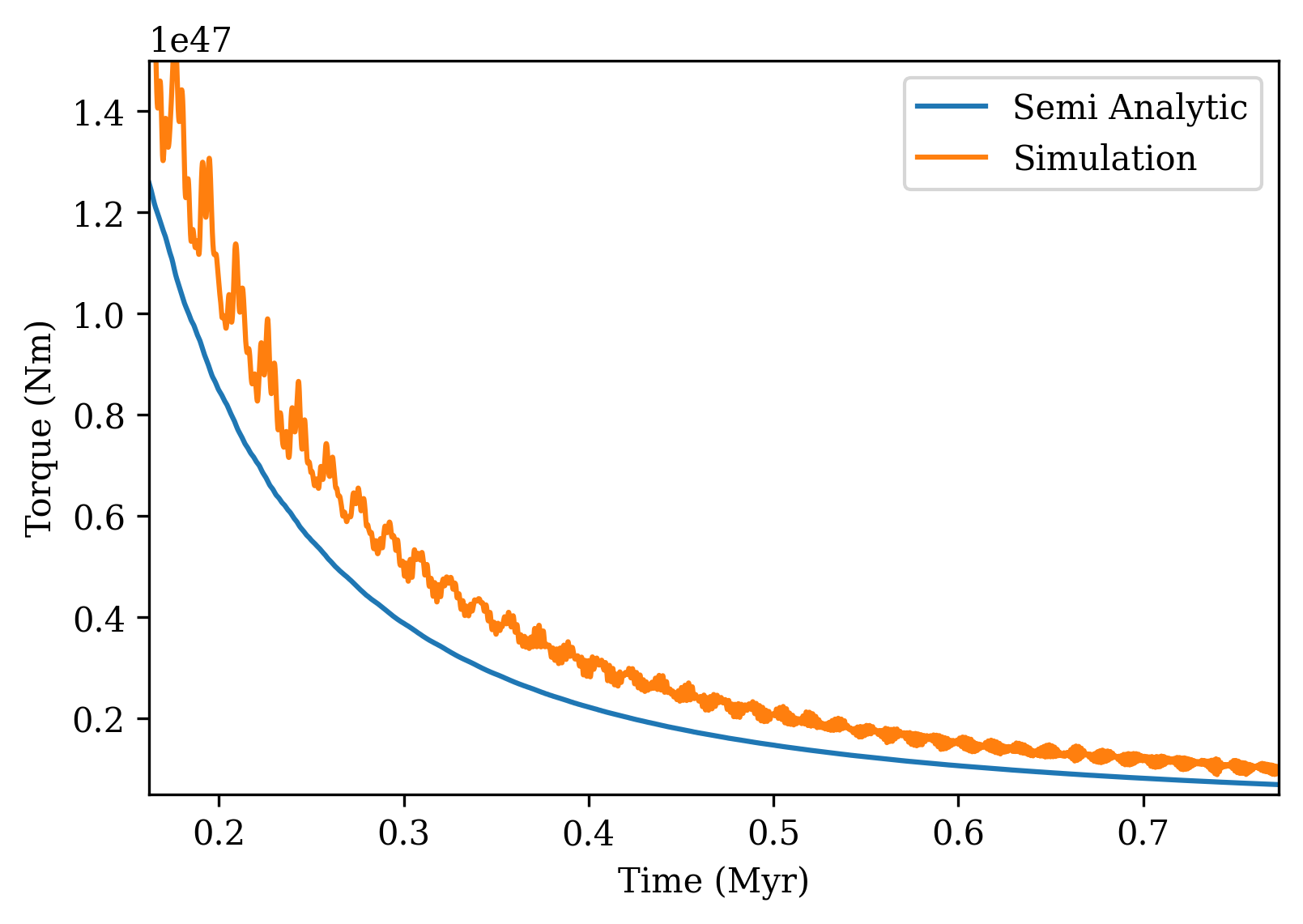}
    \caption{Comparisons of the torque on the black holes in the semi-analytic model, \ref{eq:torque-simp}, to our smoothed simulations, using a running average of 2000 steps (top), 10000 steps (middle), and 25000 steps (bottom).}
    \label{fig:torque-comp}
\end{figure}

In reality, the actual black hole separation does not decrease monotonically, so the instantaneous torque on the SMBH must differ significantly from Equation~\ref{eq:torque-simp}. Figure~\ref{fig:torque-comp} plots the inferred torque at different levels of smoothing and we see a similar situation to our earlier analysis \cite{Boey:2024dks} of black holes moving radially inside a soliton -- the average motion inferred from the dynamical friction is reasonably accurate, but instantaneous predictions  differ dramatically from the simulations. The estimated torque is consistently below the calculated value, but this is offset by the underestimation of initial velocity in the semi-analytic model.

Annulli, Cardoso and Vicente \cite{Annulli:2020lyc} estimate  the drag on a binary SMBH by looking directly at its coupling to the soliton, an  arguably more physical strategy than using an empirical approximation to the dynamical friction.  They find that the drag on an equal mass binary is
\begin{equation}
    \dot{T}=-\frac{192\pi(2\pi)^{5/3}(2M)^{5/3}}{20T^{5/3}}-\frac{3.1M^4_{s}M(2M)^{2/3}T^{17/6}}{10^3m^{-17/2}},
\end{equation}
where $M_s$ is the mass of the soliton, T is the period, and $G$, $c$, and $\hbar$ are equal to unity. The first term is associated with gravitational wave emission, which is not relevant at the range of separations in our simulations, so after dropping this, restoring factors of $G$ and $\hbar$ and again making use of Kepler's third law we find
\begin{equation}
    \dot{D}=-\kappa D^{15/4}
\end{equation}
where 
\begin{equation}
    \kappa=\frac{2}{3}\frac{3.1G^{17/3}M^4_{s}M(2M)^{2/3}}{10^3(\frac{m}{\hbar c^2})^{-17/2}}\left(\frac{4\pi^2}{G(2M)}\right)^{11/12}.
\end{equation}
This integrates in a similar fashion to Equation (\ref{eq:drdt-integral}), 
\begin{equation}
    D=\left(\frac{11}{4}\kappa (t-t_0)+D_0^{-11/4}\right)^{-4/11} \, .
    \label{eq:drdt-integral-Annuli}
\end{equation}
The exponent $4/11$ is very close to the $2/5$ derived by Koo {\em et al.\/}, which provided a tolerable but not perfect fit. Unlike Equation \ref{eq:drdt-integral}, however, this expression has no free parameters. In particular, for our system  $\kappa = 100.8\textrm{pc}^{-11/4}\textrm{Myr}^{-1}$ but a numerical fit (starting 100000 timesteps into the simulation)  gives  $\kappa = 12.16\textrm{pc}^{-11/4}\textrm{Myr}^{-1}$, so the observed orbital decay is slower than this expression predicts. Nevertheless, there is no arbitrary constant in this result and Annulli {\em et al.\/} make it clear that their calculation is approximate, so this formalism and that of Koo {\em et al.} are arguably equally effective. 

\begin{figure}
    \centering
    \includegraphics[width=\linewidth]{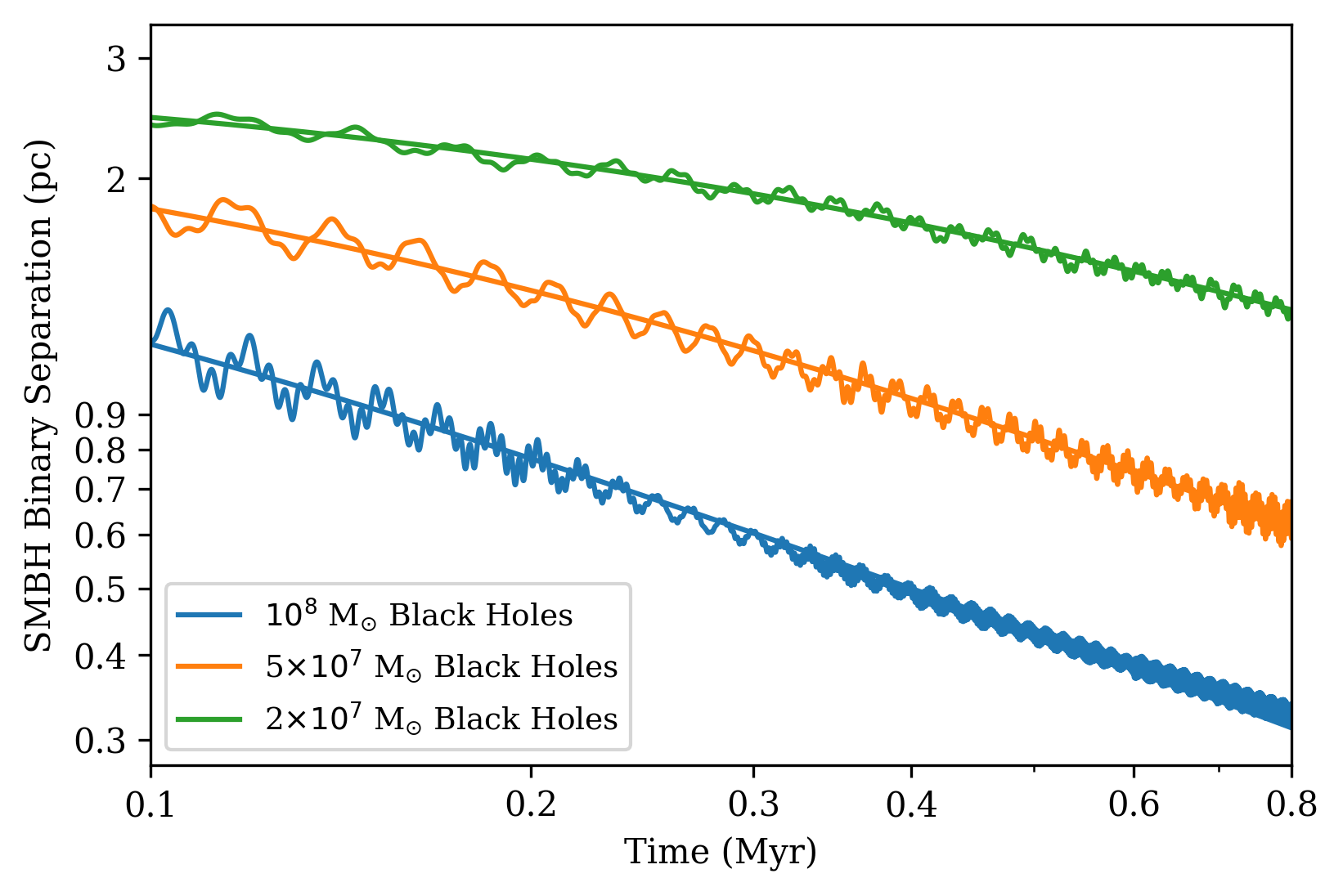}
    \caption{Fits of the empirical relationship (equation (\ref{eq:numerical-fit})) to SMBH binaries with each black hole  2, 5 and 10\%  of the mass of the soliton.}
    \label{fig:Numerical-fit}
\end{figure}

Motivated by equations (\ref{eq:drdt-integral}) and (\ref{eq:drdt-integral-Annuli}), we consider a generic empirical fit of the form
\begin{equation}
    D=A(1+Bt)^{-C} \, ,
    \label{eq:numerical-fit}
\end{equation}
corresponding to a rate of orbital decay 
\begin{equation}
    \dot{D}=-A^{-1/C}BCD^{(1+C)/C} \, ,
    \label{eq:numerical-fit-derivative}
\end{equation}
We fit coefficients $A, B, $ and $C$ to our simulations, illustrating the results in Figure \ref{fig:Numerical-fit} with the corresponding numerical coefficients shown in Table~\ref{tab:fits}. There is no clear trend in the exponent $C$ and the typical value of $\sim 0.7$  differs substantially from  the semi-analytic models, which are much closer to each other than they are to the numerical results. The $B$ parameter increases with the SMBH mass, reflecting the more rapid decay of larger systems. Conversely, $A$  is fixed by the chosen initial radius, since $D \rightarrow A$ as $t \rightarrow 0$ and in this limit the other parameters are largely irrelevant.

\begin{table}[tb]
    \centering
    \begin{tabular}{|r||c|c|c|}
\hline
\hline
Ratio & \hspace{5mm}$A \textrm{ (pc})$\hspace{5mm} & 
\hspace{5mm}$B\textrm{ (Myr}^{-1})$\hspace{5mm} & \hspace{5mm}$C$\hspace{5mm} \\
\hline\hline
2\% &      2.94&3.12& 0.662\\
5\% &  2.78 & 7.44 & 0.777 \\
10\% &     2.72 & 22.7 & 0.733 \\ 
\hline
    \end{tabular}
    \caption{Fits to equation (\ref{eq:numerical-fit}) for binary SMBH systems.}
    \label{tab:fits}
\end{table}

\begin{figure}[tb]
    \centering
    \includegraphics[width=\linewidth]{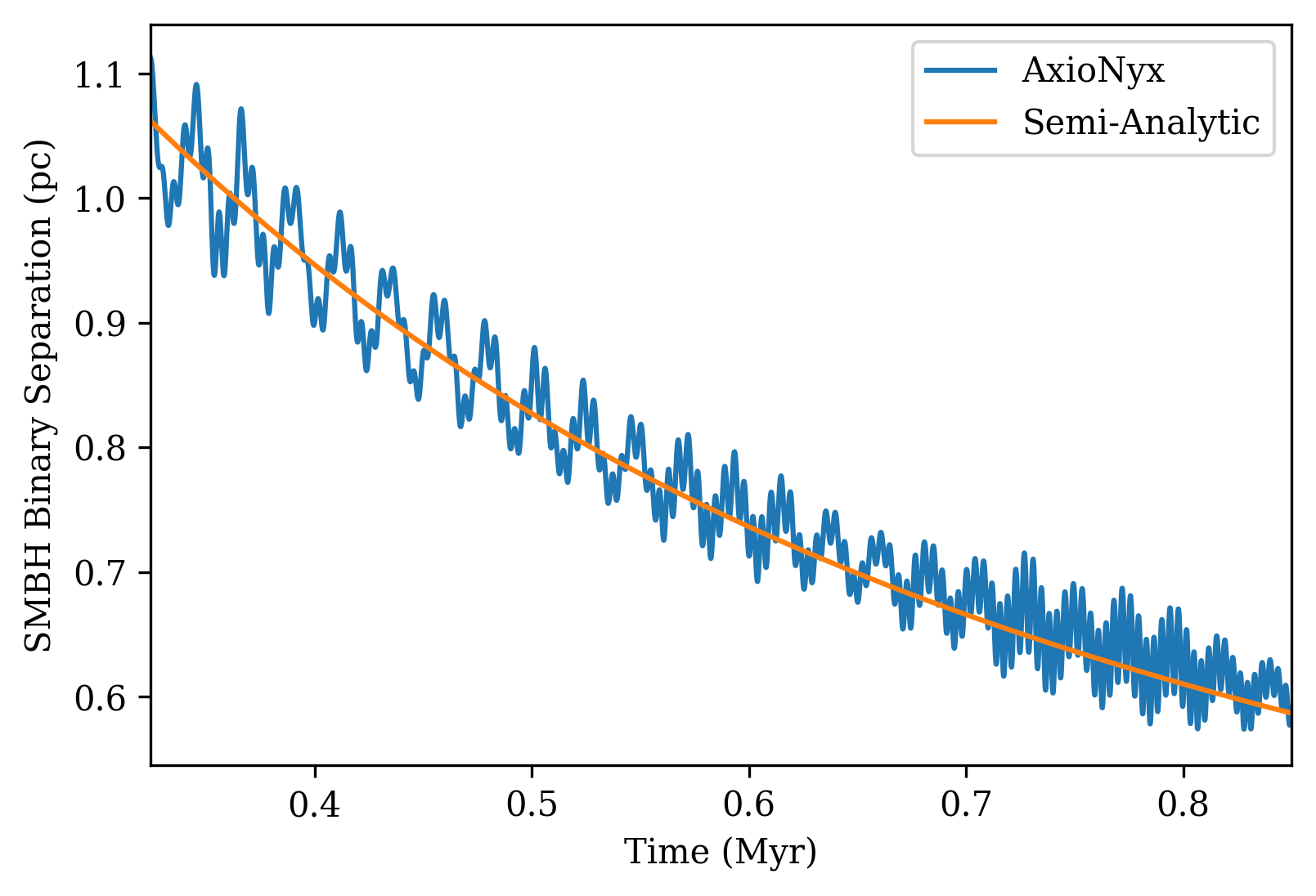}
    \caption{A numerical integration of the semi-analytic model (equation \ref{eq:rdotfull}) with the ``pinched'' $\rho(r)$ is  plotted alongside results from {\sc AxioNyx} with $5\times 10^7 \mathrm{M}_\odot$ black holes, starting 100000 timesteps into the simulation to reduce  initial conditions dependence.}
    \label{fig:Numerical_integration_fit}
\end{figure}

By default, equation (\ref{eq:rdotfull}) is integrated with the unperturbed soliton profile.  Substituting the time-averaged pinched profile improves the match to the full simulations; results for $5 \times 10^7 \mathrm{M}_\odot$ black holes  are shown in Figure~\ref{fig:Numerical_integration_fit}. In this case the value of $C$ derived from the semi-analytic treatment  ($C=0.807$) is close to that obtained from the numerical simulation. 

The difference appears to stem from the fact that the ``pinched'' profile is still relatively steep when $r\sim 0.5$pc, causing the drag to decrease more slowly than otherwise expected as $r$ decreases, boosting the effective value of $C$. However, when $r$ is very small and $\rho(r)$  is close to its maximum the effective value of $C$ could return to  $0.4$ but our simulations do not fully explore this regime. That said, we do see some evidence for a ``turnover'' if we fit only to the very last stages of the fiducial run and this remains a topic for future work.

\section{The Final Parsec and Nanohertz Gravitational Waves}
\label{sec:parsec}

The separation of an equal-mass binary system in a circular orbit decays via gravitational wave emission at a rate \cite{Peters:1964zz}
\begin{equation}
   \dot{D}_{\textrm{GW}}=-\frac{128}{5}\frac{G^3 M^3}{D^3},
    \label{eq:gw-decay}
\end{equation}
where $D=(2GM)^{1/3}/(\pi f_{\textrm{r}})^{2/3}$, $f_{r}$ is the source-frame gravitational wave frequency and we assume $M_{enc} \ll M$. A binary with a starting separation of $D_0$ will thus merge after time\footnote{In this section we consider only equal-mass binaries, and thus make use of the black hole mass $M$ rather than the chirp mass $\mathcal{M}$, noting that the expressions can be re-framed in terms of the chirp mass using $M = 2^{1/5}\mathcal{M}$.}
\begin{equation}
   \tau_{\textrm{GW}}(D_0) = \frac{5}{512} \frac{ {D_0}^4}{G^3 M^3 } \, ,
\end{equation}
due to gravitational wave emission alone. Orbital decay is thus driven first by dynamical friction and later by gravitational wave emission. If ULDM dynamical friction can reduce the separation from a few parsecs to the point where  gravitational wave emission alone will result in coalescence within $10^{10}$yr we can assume that a merger is guaranteed within the current lifetime of the universe. For $10^8\textrm{M}_{\odot}$ black holes, $\tau_{\textrm{GW}} < 10^{10}$yr if $D_0 \le 0.076$pc. 

Using the parameters from Table \ref{tab:fits} we calculate that dynamical friction can drive our fiducial binary from a separation of 1pc to 0.076pc in the relatively short time of 5.6Myr. For  $2 \times 10^7\textrm{M}_{\odot}$ black holes $\tau_{\textrm{GW}} < 10^{10}$yr if $D_0 \le 0.023$pc and our results suggest that dynamical friction will achieve this separation in $477$Myr,  easily less than a Hubble time. Consequently, at least for relatively large black holes in ULDM solitons,  dynamical friction does appear to ameliorate the final parsec problem. 

The natural question that follows is whether this drag will still be significant for SMBH binaries in the pulsar timing band. The relationship between separation and frequency is 
\begin{equation}
    D=\left(\frac{2GM}{\pi^2f_{gw}^2}\right)^{\frac{1}{3}},
\end{equation}
where we have assumed that the soliton mass enclosed is negligible at these separations. The gravitational wave frequency, $f_{gw}$, is twice the orbital frequency. 

\begin{figure}[tb]
    \centering
    \includegraphics[width=\linewidth]{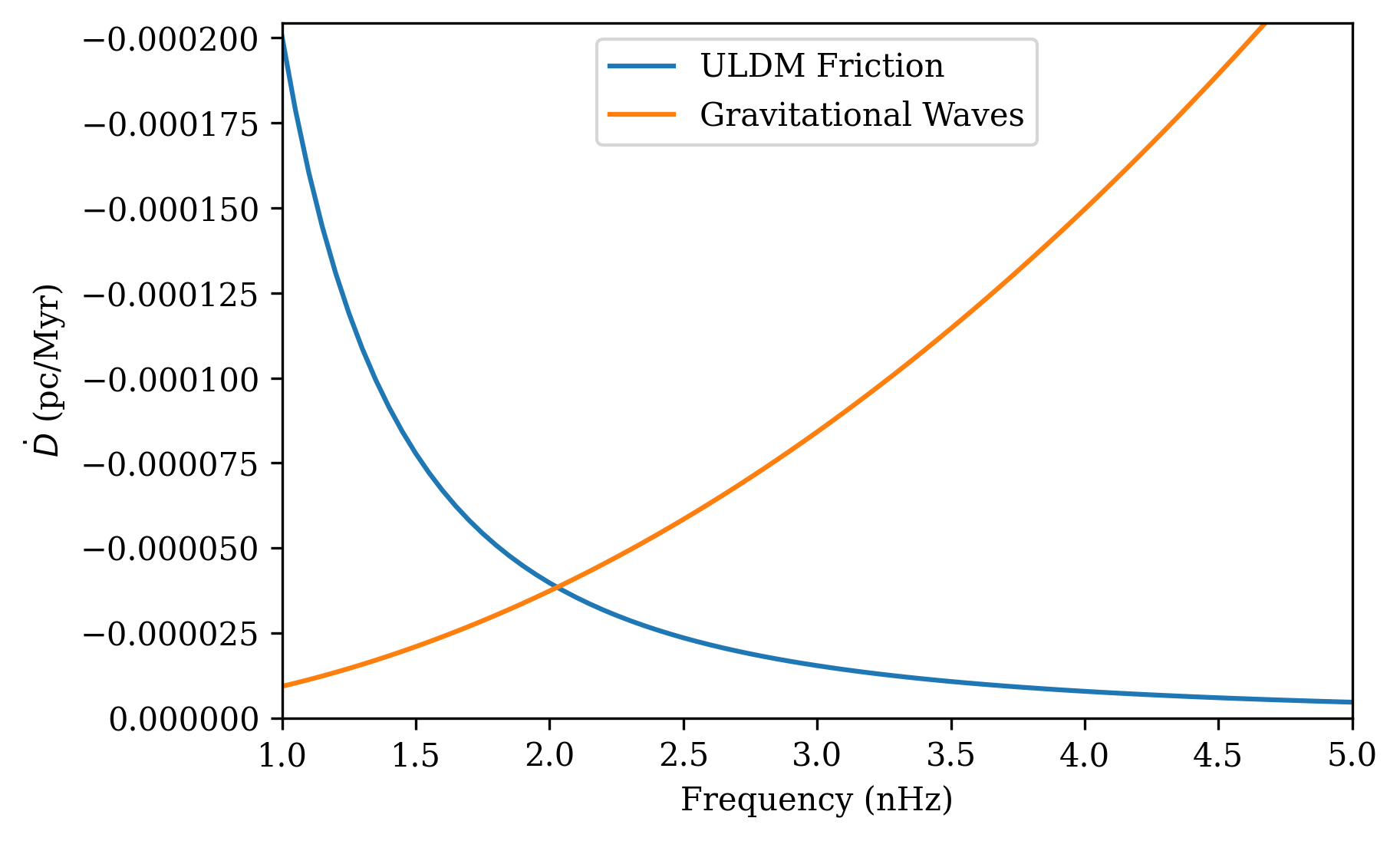}
    \caption{The separation decay induced by ULDM drag and gravitational waves for our fiducial system.}
    \label{fig:Drag_GW_Fiducial}
\end{figure}

As discussed at the end of Section~\ref{sec:semi-analytic} in this regime equation (\ref{eq:drdt-integral}) may yield a better estimate of the drag than (\ref{eq:numerical-fit-derivative});  we set $C=0.4$ and  use equation~(\ref{eq:drdt-simp}) to characterize the drag in what follows. Comparing this contribution to equation~(\ref{eq:gw-decay}) we can quantify the two contributions to $\dot D$;  results for the fiducial parameters with a fit starting at 130000 steps and a corresponding $\alpha=0.341$ are shown in Figure \ref{fig:Drag_GW_Fiducial}. Dynamical friction dominates in the lowest parts of the PTA band,  reducing the gravitational wave flux from this system at very low frequencies.

\begin{figure}[tb]
    \centering
    \includegraphics[width=1\linewidth]{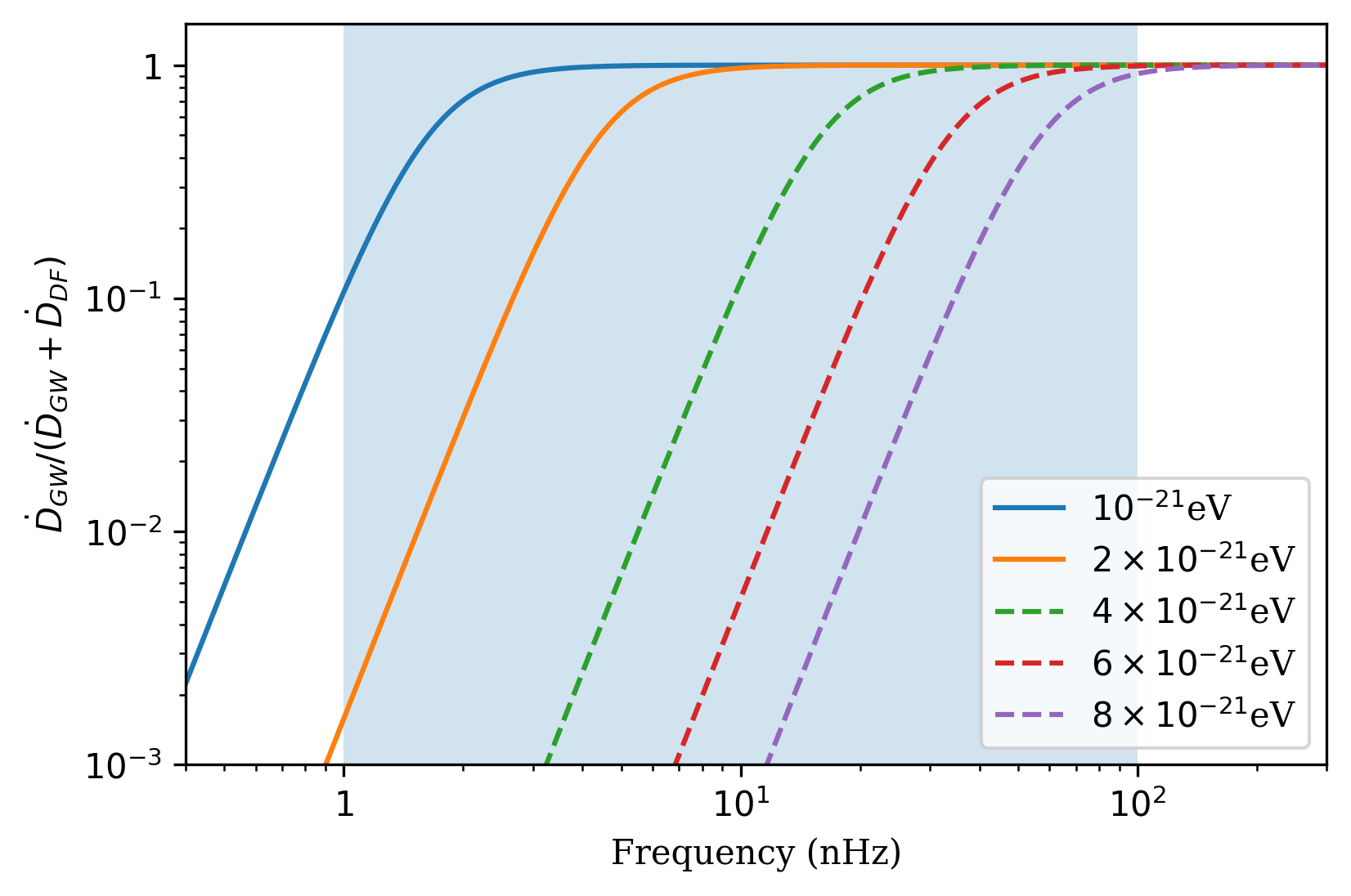}
    \caption{Ratio of the decay due to gravitational wave emission to the sum of the decay modes. The two solid lines extrapolate fits to  simulations while the dashed lines are estimated from equation (\ref{eq:K}). The PTA frequency range is indicated by the shaded region.}
    \label{fig:particle_mass_ratios}
\end{figure}

Dynamical friction rises rapidly with ULDM particle mass when the other parameters are held fixed.   Figure \ref{fig:particle_mass_ratios} shows estimates of the fractional contribution to orbital decay from gravitational wave emission for $5\times 10^7 \mathrm{M}_\odot$ black holes and a range of ULDM masses.  The two solid lines correspond to extrapolations of  simulations shown in  Figure~\ref{fig:ULDM_Particle_Mass_Dependence}. The dashed lines for higher ULDM masses are estimates generated by assuming the solitons have not undergone any pinching, taking $\alpha=0.208$ from a fit to the simulation with a $2\times10^{-21}$~eV ULDM mass\footnote{This simulation ends with the $v^3$ and $4\pi r^2\rho Gv$ terms in equation~(\ref{eq:rdotfull}) of similar order so this value of $\alpha$ necessarily represents an extrapolation.}, and calculating $K$ from equation (\ref{eq:K}). These results suggest that ULDM solitons could significantly depress gravitational emission in the pulsar timing band and would do so more noticeably at lower frequencies. 

Intriguingly, this effect could improve the match to the observed spectrum, which lacks power at the lower end of the band relative to naive expectations \cite{NANOGrav:2023gor,Ellis:2023dgf}.   That said, we also note that the core-halo relation itself depends on particle mass, with the soliton fraction decreasing with $m$. The soliton mass is held constant throughout, so the estimates with larger $m$ implicit apply to larger halo - although we also expect larger expect larger black holes in these systems, and the SMBH is also held fixed in this analysis. This complements an analysis of PTA timing residuals by Tomaselli \cite{Tomaselli:2024ojz}, which also finds a possible suppression due to ULDM drag at lower frequencies.

\section{\label{sec:conclusion}Conclusion and Discussion}

We have performed high resolution, non-relativistic simulations of binary SMBH moving in a ULDM soliton with the adaptive mesh refinement code {\sc AxioNyx}, allowing these simulations to run over longer durations and at higher resolutions than in previous analyses. We examined the sensitivity to the black hole mass, the soliton mass and the ULDM particle mass.  We have focused on solitons that would be typical of galactic halos hosting SMBH significantly larger than their Milky Way analogues. However, these systems are likely to dominate the stochastic background generated by SMBH mergers and are thus the most physically relevant. 

Our results give insight into semi-analytic approximations to the dynamical friction experienced by black holes in ULDM \cite{Hui:2016ltb}.  As was the case with radial passage through the soliton the detailed forces experienced by the black holes in our simulations are very different from those predicted by a semi-analytic approximation \cite{Boey:2024dks}, but the time-averaged force may still be well-modeled.  That said, semi-analytic approximations need to be used with care as the presence of an SMBH at the center of a soliton changes its profile, raising the central density and causing $\rho(r)$ to vary more rapidly with $r$. As a consequence, we find that the approach to the center occurs more rapidly than suggested by both simple estimates and previous numerical studies \cite{Koo:2023gfm}. This highlights the importance of realistic simulations; these are facilitated by the use of AMR-based codes \cite{Schwabe:2020eac} whose efficiency relative to fixed-grid codes (e.g. \cite{Edwards:2018ccc}) justifies their greater complexity. Further development to the semi-analytic model may improve this mismatch; in particular, implementing a soliton profile that accounts for the potential from the binary is a possible avenue for improvement.

A single black hole orbiting inside the soliton can undergo ``stone-skipping'', with  significant re-excitement of its orbit after an initial period of decay \cite{Wang:2021udl}. By contrast, in binary systems the  separation decreases steadily, with an overlaid  modulation driven by a breathing mode in the soliton which is itself excited by the moving black holes.   We find that large black holes in solitons at the centers of large halos undergo rapid orbital decay even at sub-parsec separations,  particularly with relatively large ULDM particle masses. Consequently, it is plausible that ULDM solitons could contribute to solving the final parsec problem in large halos. 

Extrapolating our results to  orbital separations associated with gravitational wave emission in the pulsar timing band it also appears possible that the drag can be nontrivial in this regime, particularly at lower frequencies. As with decay through the final parsec the drag  depends strongly on the halo mass which is correlated with the expected sizes of both the soliton and the embedded black holes. Given that the SMBH background in the pulsar timing band is thought to be dominated by massive systems \cite{Ellis:2023owy} this raises the possibility that ULDM could reduce the amplitude of this background. 

Intriguingly, any suppression would be frequency dependent, so could potentially account for the shape of the observed PTA spectrum \cite{NANOGrav:2023gor}. Moreover,  the relatively strong suppression seen with larger ULDM particle masses may even provide a new approach to constraining this parameter. That said, any such analysis has a number of complexities. Firstly, the core:halo relation is itself subject to some uncertainty \cite{Schive:2014hza,Schwabe:2016rze,Mocz:2017wlg, Padilla:2020sjy,Burkert:2020laq,Nori:2020jzx,Mina:2020eik,Taruya:2022zmt,Zagorac:2022xic,Kendall:2023kit} which will change the central density for a given halo, and the solitonic  core resulting from a recent merger is unlikely to be relaxed. Secondly, our discussion has assumed that a single ULDM species accounts for all dark matter, but the overall parameter space faces a number of constraints \cite{Schive:2015kza,Corasaniti:2016epp,Menci:2017nsr,Ni:2019qfa,Winch:2024mrt,Irsic:2017yje,Rogers:2020ltq,Schutz:2020jox,DES:2020fxi,Banik:2019smi,Sipple:2024svt}.  However, many of these limits relax if only a fraction of the dark matter consists of ULDM, or if there are multiple ULDM species. In the former case the soliton survives if the ULDM fraction is greater than roughly 10\% \cite{Schwabe:2020eac} but will be less massive and less dense, reducing any damping. In the latter case, nested solitons can form \cite{Gosenca:2023yjc,Luu:2024gnk,Chen:2023unc,Jain:2023ojg}, complicating the dynamics and reducing the drag relative to a scenario in which all of the dark matter was accounted for by the most massive ULDM species present. 

Our simulations are simplified by starting with the SMBH binary moving inside a stationary soliton. Moreover, by  omitting the wider halo we  suppress the random walk of the soliton position driven by interactions with ULDM granules in the halo \cite{Schive:2019rrw,Chowdhury:2021zik}, which will could reheat the SMBH binary. Consequently, it would be interesting (if computationally expensive) to simulate the full halo merger while accounting for the SMBH dynamics. Moreover, these simulations have focused on equal mass binaries but the single black hole case suggests that significantly unequal mass binaries may behave quite differently, potentially experiencing stone-skipping effects at large mass ratios. These possibilities all provide interesting directions for future investigation.

This work also overlaps with a range of investigations in related domains. The impact of UDLM at the late stages of black hole mergers has been studied in numerical relativity \cite{Bamber:2022pbs,Aurrekoetxea:2023jwk}, and have similarly found increased orbital decay and adjustments to predicted gravitational wave signals, and further work is needed to connect these two regimes. 
Singh {\em et al.\/} \cite{Singh:2025uvp} have studied the effects of ULDM solitons on gravitational wave strain in the LISA band due to lensing, finding adjustments to the amplification factor and time-domain integral, a complement to the work we have performed for the PTA band.
Separately, there is a possibility of black hole formation \cite{Eggemeier:2021smj} in a long matter dominated phase in the post-inflationary universe \cite{Amin:2019ums,Musoke:2019ima,Niemeyer:2019gab,Eggemeier:2020zeg}. Their potential interactions with the remnant inflaton, which is described by the  Schr\"{o}dinger-Poisson equation, and one another are currently unexplored. 

In summary, this paper marks a significant step forward in analyses of SMBH-soliton interactions and the wider study of ULDM dynamics.  We have seen that dynamical friction from the soliton could help to resolve the final parsec problem and potentially modify gravitational wave production at nanohertz frequencies. 

\begin{acknowledgments}
We thank Chris Gordon, Peter Hayman, Hyeonmo Koo,  Jens Niemeyer, Shreyas Tiruvaskar and Luna Zagorac for
useful conversations during the course of this work. This work is
supported by the Marsden Fund of the Royal Society of New Zealand and we acknowledge the use of New Zealand eScience Infrastructure (NeSI) high performance computing facilities.
\end{acknowledgments}

\appendix


\bibliography{apssamp}

\end{document}